\documentclass[12pt]{article}
\usepackage{amsmath}
\usepackage[unicode,bookmarks,bookmarksopen,bookmarksopenlevel=2,colorlinks,linkcolor=red,citecolor=green]{hyperref}

\def\comment#1{}
\input{tcilatex}

\begin{document}

\title{Explicit solutions of the WDVV equation determined by the ``flat''
hydrodynamic reductions of the Egorov hydrodynamic chains.}
\author{Maxim V. Pavlov \\
Lebedev Physical Institute, Moscow}
\date{}
\maketitle

\begin{abstract}
Classification of the Egorov hydrodynamic chain and corresponding 2+1
quasilinear system is given in \textbf{\cite{Maks+Egor}}. In this paper we
present a general construction of explicit solutions for the WDVV equation
associated with Hamiltonian hydrodynamic reductions of these Egorov
hydrodynamic chains.
\end{abstract}

\tableofcontents

\bigskip \bigskip

\textit{keywords}: hydrodynamic chain, Hamiltonian structure, Miura type
transformation, hydrodynamic reduction, hydrodynamic type system, WDVV
equation.


\section{Introduction}

The associativity equations (or WDDV equations) appeared in the
classification problem for the topological field theories at the early 90's,
see \textbf{\cite{Dijkgraaf}} and \textbf{\cite{Dubr}}. During the last
years these equations has attracted a great interest due to connections with
the enumerative geometry (Gromov--Witten invariants \textbf{\cite{Konc}}),
quantum cohomology \textbf{\cite{Ryan}}, the Whitham theory \textbf{\cite%
{Krich}} and centroaffine geometry \textbf{\cite{Fer+centre}}.

This paper is devoted to a construction of \textbf{explicit} solutions
(expressed via elementary or well-known special functions of flat
coordinates) of the WDVV equation (see \textbf{\cite{Dubr}}) connected with
hydrodynamic reductions of the Egorov hydrodynamic chains (see \textbf{\cite%
{Maks+Egor}}). For simplicity we restrict our consideration on the class of $%
N$ component Hamiltonian hydrodynamic type systems embedded in the famous
Benney hydrodynamic chain (see \textbf{\cite{Aoyama}}, \textbf{\cite{Benney}}%
, \textbf{\cite{Gib+Tsar}}, \textbf{\cite{Krich}}, \textbf{\cite{KM}})%
\begin{equation}
A_{t}^{k}=A_{x}^{k+1}+kA^{k-1}A_{x}^{0}\text{, \ \ \ \ \ \ \ }k=0,1,2,...
\label{bm}
\end{equation}%
and in the Kupershmidt hydrodynamic chains (see \textbf{\cite{Kuper}}, 
\textbf{\cite{Maks+Kuper}})%
\begin{equation}
B_{t}^{k}=B_{x}^{k+1}+B^{0}B_{x}^{k}+(\beta k+\gamma )B^{k}B_{x}^{0}.
\label{kuper}
\end{equation}

\textbf{Definition}: \textit{The semi-Hamiltonian} (see \textbf{\cite{Tsar}}%
) \textit{hydrodynamic type system}%
\begin{equation}
r_{t}^{i}=v^{i}(\mathbf{r})r_{x}^{i}\text{, \ \ \ \ \ \ \ }i=1,2,...,N
\label{rims}
\end{equation}%
\textit{possessing the couple of conservation laws}%
\begin{equation}
a_{t}=b_{x}\text{, \ \ \ \ \ \ \ \ \ \ }b_{t}=c_{x}  \label{egor}
\end{equation}%
\textit{is said to be \textbf{Egorov} (see} \textbf{\cite{Maks+Tsar}}; 
\textit{and also} \textbf{\cite{Maks+Egor+int}}).

It means, that the conservation law density $a$ is a potential of the Egorov
diagonal metric $g_{ii}=H_{i}^{2}$, where the Lame coefficients $H_{i}$ can
be found from (see \textbf{\cite{Tsar}})%
\begin{equation}
\partial _{i}\ln H_{k}=\frac{\partial _{i}v^{k}}{v^{i}-v^{k}}\text{, \ \ \ \
\ \ \ \ \ }i\neq k.  \label{lame}
\end{equation}

Suppose some Egorov hydrodynamic type system has the local Hamiltonian
structure%
\begin{equation}
c_{t}^{i}=\partial _{x}\left( \bar{g}^{ik}\frac{\partial \mathbf{h}}{%
\partial c^{k}}\right) \text{, \ \ \ \ \ \ \ }i=1,2,...,N,  \label{hama}
\end{equation}%
where $\bar{g}^{ik}$ is a constant symmetric non-degenerate matrix (see
details in \textbf{\cite{Dubr+Nov}}). Then, a corresponding solution of the
WDVV equation can be found (see \textbf{\cite{Aoyama}}, \textbf{\cite{Dubr}}%
, \textbf{\cite{Krich}}; and also \textbf{\cite{china}}, \textbf{\cite%
{Eguchi}}). The first publication \textbf{\cite{Dubr} }describing this
connection of the Egorov hydrodynamic type systems and the WDVV equation was
appear 14 years ago, but a deficit of explicit solutions still exists.

In this paper we present an \textit{effective algorithm} allowing to
construct infinitely many particular solutions of the WDVV equation written
in an \textit{explicit form} via flat coordinates $a^{k}$ of corresponding
Egorov hydrodynamic type systems. These hydrodynamic type systems are
Hamiltonian hydrodynamic reductions (\textbf{\ref{hama}}) of the Egorov
hydrodynamic chains (see \textbf{\cite{Maks+Egor}}). Thus, in this paper we
show how these hydrodynamic reductions can be converted in solutions of the
WDVV equation.

Let us emphasize again: any solution of the WDVV equation determines an
integrable hierarchy of the Egorov hydrodynamic type systems; each Egorov
Hamiltonian hydrodynamic type system determines a solution of the WDVV
equation. It means that a description of all Egorov Hamiltonian hydrodynamic
type systems is equivalent to a description of all solutions of the WDVV
equation. Thus, in this paper we are able to present a large list of new
solutions of the WDVV equation written in an explicit form via flat
coordinates.

The \textbf{scheme}: The algorithm presented in this paper contains
following six steps:

\textbf{1}. Suppose the \textbf{Egorov} Hamiltonian hydrodynamic type system
(\textbf{\ref{hama}}) is given via the Riemann invariants (\textbf{\ref{rims}%
}). Then the Lame coefficients $H_{k}$ can be found from (\textbf{\ref{lame}}%
), and the rotation coefficients $\beta _{ik}$ of the corresponding \textit{%
conjugate} curvilinear coordinate net are given by (see \textbf{\cite%
{Darboux}})%
\begin{equation*}
\beta _{ik}=\frac{\partial _{i}H_{k}}{H_{i}}\text{, \ \ \ \ \ \ }i\neq k.
\end{equation*}

\textbf{2}. These rotation coefficients $\beta _{ik}$ satisfy the
Bianchi--Darboux--Lame--\textbf{Egorov} system%
\begin{equation}
\partial _{i}\beta _{jk}=\beta _{ji}\beta _{ik}\text{, \ \ \ \ \ }i\neq
j\neq k\text{, \ \ \ \ \ \ \ \ \ \ }\beta _{jk}=\beta _{kj}.  \label{wave}
\end{equation}%
The linear PDE system%
\begin{equation}
\partial _{i}\tilde{H}_{k}=\beta _{ik}\tilde{H}_{i}\text{, \ \ \ \ \ \ }%
i\neq k  \label{lin}
\end{equation}%
has a general solution parameterized by $N$ arbitrary functions of a single
variable.

The existence of the local Hamiltonian structure (\textbf{\ref{hama}}) is
equivalent the zero curvature condition (see \textbf{\cite{Darboux}})%
\begin{equation}
\partial _{i}\beta _{ik}+\partial _{k}\beta _{ki}+\underset{m\neq i,k}{\sum }%
\beta _{mi}\beta _{mk}=0,  \label{flat}
\end{equation}%
which is compatible with (\textbf{\ref{wave}}). Thus, the rotation
coefficients $\beta _{ik}$ describe corresponding \textit{orthogonal}
curvilinear coordinate net.

It means that the linear PDE system (\textbf{\ref{lin}}) can be replaced by $%
N$ ODE systems (see \textbf{\cite{Tsar}})%
\begin{equation*}
\partial _{i}H_{k}^{(n,s)}=\beta _{ik}H_{i}^{(n,s)}\text{,\ \ }i\neq k\text{%
; \ \ \ \ \ \ }\partial _{i}H_{i}^{(n,s)}+\underset{m\neq i}{\sum }\beta
_{mi}H_{m}^{(n)}=H_{m}^{(n-1,s)}\text{, \ \ }n=0,1,...,
\end{equation*}%
where $H_{m}^{(0,s)}$ are solutions of $N$ ODE systems%
\begin{equation}
\partial _{i}H_{k}^{(0,s)}=\beta _{ik}H_{i}^{(0,s)}\text{, \ \ }i\neq k\text{%
; \ \ \ \ \ \ }\partial _{i}H_{i}^{(0,s)}+\underset{m\neq i}{\sum }\beta
_{mi}H_{m}^{(0,s)}=0\text{, \ \ }i,k,s=1,2,...,N.  \label{nul}
\end{equation}

\textbf{Definition}: \textit{The Lame coefficients }$\bar{H}_{k}^{(s)}\equiv
H_{k}^{(0,s)}$ \textit{are said to be \textbf{canonical} Lame coefficients}.

\textbf{3}. Let us introduce the \textbf{constant} matrix%
\begin{equation*}
\bar{g}^{ik}=\sum \bar{H}_{m}^{(i)}\bar{H}_{m}^{(k)}
\end{equation*}%
and the canonical Lame coefficients with sub-indexes $\bar{H}_{(s)k}=\bar{g}%
_{sm}\bar{H}_{k}^{(m)}$.

Let us choose any solution of (\textbf{\ref{nul}}) as $\bar{H}_{(1)k}$. Then
let us construct $N-1$ commuting flows%
\begin{equation}
r_{t^{k}}^{i}=\frac{\bar{H}_{(k)i}}{\bar{H}_{(1)i}}r_{t^{1}}^{i}\text{, \ \
\ \ \ }i=1,2,...,N\text{, \ \ \ \ \ }k=2,3,...,N,  \label{ric}
\end{equation}%
where all Riemann invariants $r^{k}$ are functions of $t^{1},t^{2},...,t^{N}$%
.

\textbf{4}. Let us introduce $N(N+1)/2$ functions $a_{kn}(\mathbf{r})$ such
that%
\begin{equation}
\partial _{i}a_{kn}=\bar{H}_{(k)i}\bar{H}_{(n)i}.  \label{sym}
\end{equation}

\textbf{Definition}: \textit{The functions }$a_{k1}$ \textit{are said to be
the \textbf{adjoint} flat coordinates }$a_{k}$.

Then all other functions $a_{kn}$ can be expressed via these field variables 
$a_{k}$, because they are connected with the Riemann invariants $r^{n}$ by
an invertible transformation $a_{k}(\mathbf{r})$. One can differentiate
adjoint flat coordinates $a_{i}$ with respect to times $t^{k}$. Then the
hydrodynamic type systems (\textbf{\ref{ric}}) written via the Riemann
invariants $r^{n}$ can be written in the symmetric (see (\textbf{\ref{sym}}%
)) conservative form%
\begin{equation}
\partial _{t^{k}}a_{i}=\partial _{t^{1}}a_{ik}(\mathbf{a})\text{, \ \ \ \ \
\ }i,k=1,2,...,N.  \label{set}
\end{equation}

\textbf{Definition \cite{Maks+Egor+int}}: \textit{The hydrodynamic type
systems} (\textbf{\ref{set}}) \textit{are said to be the \textbf{canonical
Egorov basic set}}.

\textbf{Remark}: Since the functions $a_{ik}(\mathbf{a})$ are symmetric (see
(\textbf{\ref{sym}})), then the unique function $\Omega
(t^{1},t^{2},...,t^{N})$ can be introduced and determined by its second
derivatives%
\begin{equation}
a_{ik}(\mathbf{a})=\frac{\partial ^{2}\Omega }{\partial t^{i}\partial t^{k}}%
\text{, \ \ \ \ \ \ }i,k=1,2,...,N.  \label{tau}
\end{equation}%
Thus, the hydrodynamic type systems (\textbf{\ref{set}}) can be replaced by $%
N(N-1)/2$ algebraic equations%
\begin{equation}
\frac{\partial ^{2}\Omega }{\partial t^{i}\partial t^{k}}=a_{ik}\left( \frac{%
\partial ^{2}\Omega }{\partial t^{j}\partial t^{1}}\right) \text{, \ \ \ \ \
\ }i,k=2,...,N.  \label{simple}
\end{equation}

\textbf{5}. Since the hydrodynamic type system (\textbf{\ref{rims}})
possesses the Hamiltonian structure (\textbf{\ref{hama}}), then its any
commuting flow%
\begin{equation*}
r_{\tau }^{i}=w^{i}(\mathbf{r})r_{x}^{i}\text{, \ \ \ \ \ \ \ }i=1,2,...,N
\end{equation*}%
possesses the same Hamiltonian structure%
\begin{equation*}
c_{\tau }^{i}=\partial _{x}\left( \bar{g}^{is}\frac{\partial \mathbf{\bar{h}}%
}{\partial c^{s}}\right) \text{, \ \ \ \ \ \ \ }i,k=1,2,...,N.
\end{equation*}%
Since the hydrodynamic type system (\textbf{\ref{rims}}) possesses the
Egorov pair (\textbf{\ref{egor}}), then its any commuting flow possesses the
similar Egorov pair (see \textbf{\cite{Maks+Tsar}})%
\begin{equation}
a_{\tau }=h_{x}\text{, \ \ \ \ \ \ }h_{\tau }=q_{x}.  \label{dop}
\end{equation}

\textbf{Lemma}: $N$ \textit{auxiliary commuting flows} (cf. (\textbf{\ref%
{ric}}))%
\begin{equation}
r_{t^{k}}^{i}=\frac{\bar{H}_{(k)i}}{H_{i}}r_{x}^{i}\text{, \ \ \ \ \ \ \ }%
i,k=1,2,...,N  \label{thrid}
\end{equation}%
\textit{are determined by the extra conservation law }(\textbf{\ref{egor}})%
\begin{equation}
a_{t^{k}}=\partial _{x}c_{k},  \label{ek}
\end{equation}%
\textit{where }$a$ \textit{is a potential of the Egorov metric}.

\textbf{Proof}: Indeed, if (\textbf{\ref{ek}}) is the conservation law of (%
\textbf{\ref{thrid}}), then $\partial _{i}c_{k}=\bar{H}_{(k)i}H_{i}$. Thus,
the adjoint flat coordinates $c_{k}$ determine auxiliary $N$ commuting flows
(\textbf{\ref{thrid}}).

Let us rewrite these $N$ hydrodynamic type systems in the potential form%
\begin{equation}
d\xi ^{i}=c^{i}dx+\bar{g}^{is}\frac{\partial \mathbf{h}_{k}}{\partial c^{s}}%
dt^{k}.  \label{dif}
\end{equation}%
Since the local Hamiltonian structure preserves under arbitrary linear
change of independent variables ($x,t^{k}$), then one can introduce a new
set of flat coordinates (see \textbf{\cite{Maks+notice}})%
\begin{equation}
a^{k}=\bar{g}^{ks}\frac{\partial \mathbf{h}_{1}}{\partial c^{s}},
\label{flet}
\end{equation}%
then (\textbf{\ref{dif}})%
\begin{equation}
d\tilde{\xi}^{i}=\bar{g}^{is}\frac{\partial \mathbf{h}_{1}}{\partial c^{s}}%
dt^{1}+\bar{g}^{is}\frac{\partial \mathbf{h}_{k}}{\partial c^{s}}%
dt^{k}=a^{i}dt^{1}+\bar{g}^{is}\frac{\partial \mathbf{\tilde{h}}_{k}}{%
\partial a^{s}}dt^{k}  \label{difa}
\end{equation}%
leads to the

\textbf{Lemma \cite{Dubr}}: \textit{The Hamiltonian densities} $\mathbf{%
\tilde{h}}_{k}$ \textit{are determined by the unique function} $%
F(a^{1},a^{2},...,a^{N})$%
\begin{equation}
\mathbf{\tilde{h}}_{k}=\frac{\partial F}{\partial a^{k}}.  \label{der}
\end{equation}

\textbf{Proof}: $N-1$ commuting flows (\textbf{\ref{set}}) possess the local
Hamiltonian structure (see (\textbf{\ref{difa}}))%
\begin{equation}
\partial _{t^{k}}(\bar{g}_{ns}a^{s})=\partial _{t^{1}}\frac{\partial \mathbf{%
\tilde{h}}_{k}}{\partial a^{n}}.  \label{zim}
\end{equation}%
If to substitute $k=1$ in (\textbf{\ref{zim}}), then%
\begin{equation}
\bar{g}_{is}a^{s}=\frac{\partial \mathbf{\tilde{h}}_{1}}{\partial a^{i}}.
\label{e}
\end{equation}%
If to substitute $n=1$ in (\textbf{\ref{zim}}) and to take into account that
by definition $a_{i1}\equiv a_{i}$ (see (\textbf{\ref{sym}})), then%
\begin{equation}
\frac{\partial \mathbf{\tilde{h}}_{k}}{\partial a^{1}}=a_{k}.  \label{h}
\end{equation}%
The comparison (\textbf{\ref{h}}) with (\textbf{\ref{e}}) implies the
existence of some potential function $F(a^{1},a^{2},...,a^{N})$ (see (%
\textbf{\ref{der}})).

\textbf{6}. Thus, these $N-1$ commuting flows (\textbf{\ref{zim}}) can be
written in the form%
\begin{equation}
a_{t^{k}}^{i}=\partial _{t^{1}}\left( \bar{g}^{is}\frac{\partial ^{2}F}{%
\partial a^{s}\partial a^{k}}\right) .  \label{zima}
\end{equation}

\textbf{Corollary \cite{Dubr}}: Since (see (\textbf{\ref{der}}), (\textbf{%
\ref{e}}) and (\textbf{\ref{h}}))%
\begin{equation*}
\bar{g}_{is}a^{s}=\frac{\partial ^{2}F}{\partial a^{i}\partial a^{1}},
\end{equation*}%
then (with the aid of a differentiation)%
\begin{equation*}
\frac{\partial ^{3}F}{\partial a^{i}\partial a^{k}\partial a^{1}}=\bar{g}%
_{ik}
\end{equation*}%
and (with the aid of an integration)%
\begin{equation}
\mathbf{\tilde{h}}_{1}=\frac{\partial F}{\partial a^{1}}=\frac{1}{2}\bar{g}%
_{sk}a^{s}a^{k}.  \label{mom}
\end{equation}%
This is the first nontrivial conservation law density (except the flat
coordinates $a^{k}$ and the Hamiltonian densities $\mathbf{\tilde{h}}_{k}$ ($%
k=2,3,...,N$). This is a momentum density, and a corresponding conservation
law is given by%
\begin{equation*}
\partial _{t^{k}}\mathbf{\tilde{h}}_{1}=\partial _{t^{1}}\left( a^{s}\frac{%
\partial \mathbf{\tilde{h}}_{k}}{\partial a^{s}}-\mathbf{\tilde{h}}%
_{k}\right) .
\end{equation*}

\textbf{Corollary \cite{Dubr}}: In general case $\bar{g}_{11}\neq 0$, then
(see (\textbf{\ref{mom}})) the function $F$ is \textit{cubic} with respect
to $a^{1}$, i.e.%
\begin{equation*}
F=\frac{1}{6}\bar{g}_{11}(a^{1})^{3}+\frac{1}{2}\underset{k>1}{\sum }\bar{g}%
_{1k}a^{k}(a^{1})^{2}+\frac{1}{2}\underset{k,s>1}{\sum }\bar{g}%
_{sk}a^{s}a^{k}a^{1}+\Psi (a^{2},a^{3},...,a^{N}),
\end{equation*}%
where the function $\Psi (a^{2},a^{3},...,a^{N})$ can be determined from the
compatibility conditions%
\begin{equation}
\partial _{t^{k}}(\partial _{t^{n}}a_{i})=\partial _{t^{n}}(\partial
_{t^{k}}a_{i})\text{, \ \ \ \ \ }i,k,n=2,3,...,N\text{.}  \label{cross}
\end{equation}

\textbf{Theorem \cite{Dubr}}: \textit{The above function} $%
F(a^{1},a^{2},...,a^{N})$ \textit{satisfies the WDVV equation}%
\begin{equation}
\frac{\partial ^{3}F}{\partial a^{k}\partial a^{i}\partial a^{s}}\bar{g}^{sp}%
\frac{\partial ^{3}F}{\partial a^{p}\partial a^{j}\partial a^{n}}=\frac{%
\partial ^{3}F}{\partial a^{j}\partial a^{i}\partial a^{s}}\bar{g}^{sp}\frac{%
\partial ^{3}F}{\partial a^{p}\partial a^{k}\partial a^{n}}.  \label{wdvv}
\end{equation}

\textbf{Proof}: The substitution (\textbf{\ref{zima}}) in (\textbf{\ref%
{cross}}) yields the WDVV equation.

\textbf{Observation}: Let us compare (\textbf{\ref{set}}), (\textbf{\ref{tau}%
}) and (\textbf{\ref{zima}}). Then we obtain

\begin{equation}
\frac{\partial ^{2}\Omega }{\partial t^{k}\partial t^{n}}=\frac{\partial
^{2}F}{\partial a^{k}\partial a^{n}}.  \label{ekviv}
\end{equation}%
Thus, \textit{any} $N$ \textit{component hydrodynamic reduction} (\textbf{%
\ref{set}}) \textit{of multi-dimensional nonlinear equation (cf}. (\textbf{%
\ref{simple}}))%
\begin{equation}
Q\left( \frac{\partial ^{2}\Omega }{\partial t^{k}\partial t^{n}}\right) =0
\label{zet}
\end{equation}%
\textit{can be interpreted as some solution of the WDVV equation} (\textbf{%
\ref{wdvv}}). For instance, the dispersionless limit of the KP equation
(known also as Khohlov--Zabolotzkaya equation)%
\begin{equation*}
\frac{\partial ^{2}\Omega }{\partial t\partial t}=\frac{\partial ^{2}\Omega 
}{\partial x\partial y}-\frac{1}{2}\left( \frac{\partial ^{2}\Omega }{%
\partial x^{2}}\right) ^{2}
\end{equation*}%
possesses some solutions of three component WDVV equation (\textbf{\ref{wdvv}%
}) with the extra constraint%
\begin{equation*}
\frac{\partial ^{2}F}{\partial b\partial b}=\frac{\partial ^{2}F}{\partial
a\partial c}-\frac{1}{2}\left( \frac{\partial ^{2}F}{\partial a^{2}}\right)
^{2},
\end{equation*}%
where $a,b,c$ are flat coordinates.

The Heavenly equation (see, for instance, \textbf{\cite{Heavenly}})%
\begin{equation*}
\frac{\partial ^{2}\Omega }{\partial x\partial y}\frac{\partial ^{2}\Omega }{%
\partial z\partial t}-\frac{\partial ^{2}\Omega }{\partial x\partial z}\frac{%
\partial ^{2}\Omega }{\partial y\partial t}=1
\end{equation*}%
possesses some solutions of four component WDVV equation (\textbf{\ref{wdvv}}%
) with the extra constraint%
\begin{equation*}
\frac{\partial ^{2}F}{\partial a\partial c}\frac{\partial ^{2}F}{\partial
u\partial b}-\frac{\partial ^{2}F}{\partial a\partial u}\frac{\partial ^{2}F%
}{\partial c\partial b}=1,
\end{equation*}%
where $a,b,c,u$ are flat coordinates.

In this paper we construct solutions of the WDVV equation \textbf{avoiding
the Riemann invariants} $r^{k}$ (see (\textbf{\ref{rims}})). We start from
the Egorov hydrodynamic type system written in the conservative form (%
\textbf{\ref{hama}}) and construct the canonical Egorov basic set (\textbf{%
\ref{zima}}). Then a corresponding solution of the WDVV equation can be
found in quadratures.

Let us introduce the unique function $\Omega $ depended on infinitely many
independent variables $x^{k}$ ($k=0,\pm 1,\pm 2,\pm 3,...$). Let us
introduce new functions%
\begin{equation}
\mathbf{H}_{k}=\frac{\partial ^{2}\Omega }{\partial x^{0}\partial x^{k}}%
\text{.}  \label{sec}
\end{equation}%
Suppose all other derivatives $\partial ^{2}\Omega /\partial x^{n}\partial
x^{k}$ can be expressed via $\mathbf{H}_{k}$, i.e. (cf. (\textbf{\ref{zet}}))%
\begin{equation*}
\frac{\partial ^{2}\Omega }{\partial x^{n}\partial x^{k}}=\Phi _{kn}(\mathbf{%
H}).
\end{equation*}%
Let us introduce following rule \textbf{\cite{Maks+Egor}}: if $k,n>0$, then $%
\Phi _{kn}(\mathbf{H})=\Phi _{kn}(\mathbf{H}_{0},\mathbf{H}_{1},...,\mathbf{H%
}_{k+n})$, if $k>0$, but $n<0$, then $n=-m$ and $\Phi _{-m,k}(\mathbf{H}%
)=\Phi _{-m,k}(\mathbf{H}_{-m},...,\mathbf{H}_{-1},\mathbf{H}_{0},\mathbf{H}%
_{1},...,\mathbf{H}_{k})$, if $k,n<0$, then $\Phi _{-k,-n}(\mathbf{H})=\Phi
_{-k,-n}(\mathbf{H}_{-k-n},...,\mathbf{H}_{-1},\mathbf{H}_{0})$.

\textbf{Definition \cite{Maks+Egor}}: \textit{The family of hydrodynamic
chains}%
\begin{equation*}
\partial _{x^{n}}\mathbf{H}_{k}=\partial _{x^{0}}\Phi _{kn}(\mathbf{H})
\end{equation*}%
\textit{is said to be \textbf{Egorov} hydrodynamic chains}.

Suppose all moments $\mathbf{H}_{k}$ are functions of $N$ Riemann invariants
only (see \textbf{\cite{FerKarMax}}). These functions must be compatible
with the Egorov hydrodynamic chain for all $N$. Suppose all such functions
are found. It means, that infinitely many solutions (of a corresponding
Egorov hydrodynamic chain) parameterized by $N$ arbitrary functions of a
single variable are found too. Suppose we are able to extract all
hydrodynamic reductions (\textbf{\ref{rims}}) determined by the zero
curvature condition (\textbf{\ref{flat}}). It means, that we are able to
construct a corresponding solution of the WDVV equation. This is a goal of
this paper. Without lost of generality we restrict our consideration on
well-known integrable hydrodynamic chains (\textbf{\ref{bm}}) and (\textbf{%
\ref{kuper}}). These hydrodynamic chains possess infinitely many local
Hamiltonian structures (see \textbf{\cite{Maks+Kuper}}). Thus, infinitely
many Egorov Hamiltonian hydrodynamic reductions can be converted in
corresponding solutions of the WDVV equation.

\textbf{Remark}: We use different notation for ``hydrodynamic type system
times'' $t^{k}$ and for ``hydrodynamic chain times'' $x^{n}$ to emphasize
that in general case these independent variables \textbf{do not coincide}.

\textbf{Egorov (external) criterion \cite{Maks+Egor}}: If the hydrodynamic
chain%
\begin{equation*}
A_{t}^{k}=\underset{n=0}{\overset{k+1}{\sum }}F_{n}^{k}(\mathbf{A})A_{x}^{n}%
\text{, \ \ \ \ \ \ }k=0,1,2,...
\end{equation*}%
has infinitely many $N$ component hydrodynamic reductions parameterized by $%
N $ arbitrary functions of a single variable and the couple of conservation
laws (\textbf{\ref{egor}}), then the function $\Omega $ can be determined
via an appropriate set of conservation law densities (\textbf{\ref{sec}}).

\textbf{Example \cite{Maks+Egor+int}}: The Benney hydrodynamic type system
has infinitely many $N$ component hydrodynamic reductions parameterized by $%
N $ arbitrary functions of a single variable and the couple of conservation
laws (\textbf{\ref{egor}})%
\begin{equation}
\partial _{t}A^{0}=\partial _{x}A^{1}\text{, \ \ \ \ \ }\partial
_{t}A^{1}=\partial _{x}\left( A^{2}+\frac{(A^{0})^{2}}{2}\right) .
\label{ega}
\end{equation}%
Thus, $A^{0}$ is a potential of the Egorov metric for any hydrodynamic
reduction of the Benney hydrodynamic chain.

The paper is organized in the following order. In the second section all
``flat'' hydrodynamic reductions of the Benney--DKP moment chain are
described. In the third section a new solution of the WDVV equation
associated with the so-called waterbag reduction is constructed. In the
fourth section a link between solutions of the WDVV equation and some
degenerations of the waterbag reduction are given. In the fifth section new
solution of the WDVV equation is derived from corresponding reduction of the
modified Benney hydrodynamic chain. In the sixth section new solutions of
the WDVV equations associated with corresponding hydrodynamic reductions of
the dBKP/Veselov--Novikov hierarchy are found.

\section{Benney hydrodynamic chain}

One of the most interesting questions of the classical differential geometry
which has appeared at studying of semi-Hamiltonian systems of hydrodynamic
type is the description of the surfaces admitting non-trivial deformations
with preservation of principal directions and principal curvatures. Then the
number of essential parameters on which such deformations depend, is
actually equal to number of various local Hamiltonian structures of
corresponding system of hydrodynamic type \textbf{\cite{Fer+viniti}}. Such
local Hamiltonian structures are determined by a differential-geometric
Poisson bracket of the first order \textbf{\cite{Dubr+Nov}}. It has been
proved that $N-$component hydrodynamic type system cannot possess more than (%
$N+1$) local Hamiltonian structures of Dubrovin-Novikov type \textbf{\cite%
{MaksFer+kdv}}. Integrable hydrodynamic chains (see \textbf{\cite{Maks+eps}}%
, \textbf{\cite{Maks+Egor}}, \textbf{\cite{Fer+Dav}}) are infinitely many
component generalizations of semi-Hamiltonian hydrodynamic type systems.
Thus, formally, an existence of infinitely many local Hamiltonian structures
is not forbidden. Indeed, the Kupershmidt hydrodynamic chains possess
infinitely many local Hamiltonian structures (see \textbf{\cite{Maks+Kuper}}%
). Since the modified Benney hydrodynamic chain is a particular case of the
Kupershmidt hydrodynamic chain (\textbf{\ref{kuper}}), then the Benney
hydrodynamic chain also has infinite many local Hamiltonian structures, then
infinitely many Egorov Hamiltonian hydrodynamic reductions can be extracted.
Then corresponding solutions of the WDVV equation can be obtained in an
explicit form.

The first example of integrable hydrodynamic chains was found in the theory
describing dynamics of a finite depth fluid (see (\textbf{\ref{bm}}) and 
\textbf{\cite{Benney}}).

\textbf{Theorem \cite{Gibbons}}: \textit{A deformation of the Riemann mapping%
}%
\begin{equation}
\lambda =\mu +\frac{A^{0}}{\mu }+\frac{A^{1}}{\mu ^{2}}+\frac{A^{2}}{\mu ^{3}%
}+...  \label{rm}
\end{equation}%
\textit{describes by the Gibbons equation}%
\begin{equation}
\lambda _{t}-\mu \lambda _{x}=\frac{\partial \lambda }{\partial \mu }\left[
\mu _{t}-\partial _{x}\left( \frac{\mu ^{2}}{2}+A^{0}\right) \right] ,
\label{5}
\end{equation}%
\textit{where a dynamics of the moments} $A^{k}$ \textit{is given by} (%
\textbf{\ref{bm}}).

In this section we describe \textit{truncations} of the Riemann mapping
according to different moment decompositions $A^{k}(\mathbf{r})$ (see (%
\textbf{\ref{rims}})). Such truncations of the Riemann mapping we call the 
\textit{Riemann surfaces}. We are interested in such moment decompositions $%
A^{k}(\mathbf{r})$, that corresponding Egorov hydrodynamic type systems can
be recognized as the Hamiltonian hydrodynamic reductions (\textbf{\ref{hama}}%
).

The Gibbons equation has plenty distinguished features. We mark two of them
(see \textbf{\cite{Gibbons}}).

\textbf{1}. $\lambda =\limfunc{const}$, then $\partial \lambda /\partial \mu
\neq 0$ and the Gibbons equation reduces to the generating function of
conservation laws%
\begin{equation}
\mu _{t}=\partial _{x}\left( \frac{\mu ^{2}}{2}+A^{0}\right) .  \label{zaks}
\end{equation}%
The substitution of the inverse series (\textbf{\ref{rm}})%
\begin{equation*}
\mu =\lambda -\frac{\mathbf{H}_{0}}{\lambda }-\frac{\mathbf{H}_{1}}{\lambda
^{2}}-\frac{\mathbf{H}_{2}}{\lambda ^{3}}-...
\end{equation*}%
in (\textbf{\ref{zaks}}) yields an infinite series of the Kruskal
conservation laws%
\begin{equation}
\partial _{t}\mathbf{H}_{k}=\partial _{x}[\mathbf{H}_{k+1}-\frac{1}{2}%
\overset{k-1}{\underset{m=0}{\sum }}\mathbf{H}_{m}\mathbf{H}_{k-1-m}]\text{,
\ \ \ \ \ \ }k=0,1,2,...,  \label{2}
\end{equation}%
where conservation law densities $\mathbf{H}_{k}$ are polynomial functions
of $A^{n}$. For instance, $\mathbf{H}_{0}=A^{0}$, $\mathbf{H}_{1}=A^{1}$, $%
\mathbf{H}_{2}=A^{2}+(A^{0})^{2}$, $\mathbf{H}_{3}=A^{3}+3A^{0}A^{1}$.

\textbf{Remark}: The Benney hydrodynamic chain has an infinite series of
conservation law densities $\mathbf{H}_{k}$. Since in this paper we consider
hydrodynamic reductions of the Benney hydrodynamic chain, we shall use the
special notation $\mathbf{h}_{k}(\mathbf{r})$ for corresponding \textit{%
reduced} conservation law densities $\mathbf{H}_{k}$.

\textbf{2}. The condition $\partial \lambda /\partial \mu =0$ determines the
branch points of the Riemann surface $r^{i}=\partial \lambda /\partial \mu
|_{\mu =\mu ^{i}}$, then the Gibbons equation reduces to the hydrodynamic
type system (\textbf{\ref{rims}})%
\begin{equation}
r_{t}^{i}=\mu ^{i}(\mathbf{r})r_{x}^{i}\text{, \ \ \ \ \ \ \ }i=1,2,...,N,
\label{reda}
\end{equation}%
also can be written in the conservative form (see (\textbf{\ref{zaks}}); $%
\mu \rightarrow c^{i}$)%
\begin{equation}
c_{t}^{i}=\partial _{x}\left( \frac{(c^{i})^{2}}{2}+A^{0}(\mathbf{c})\right)
,  \label{first}
\end{equation}%
where the moment $A^{0}$ is a solution of the Gibbons--Tsarev system (see 
\textbf{\cite{Maks+flat}})%
\begin{eqnarray}
(c^{i}-c^{k})\partial _{ik}A^{0} &=&\partial _{k}A^{0}\partial _{i}\left(
\sum \partial _{n}A^{0}\right) -\partial _{i}A^{0}\partial _{k}\left( \sum
\partial _{n}A^{0}\right) \text{, \ \ }i\neq k,  \notag \\
&&  \label{egt}
\end{eqnarray}%
\begin{equation*}
(c^{i}-c^{k})\frac{\partial _{ik}A^{0}}{\partial _{i}A^{0}\partial _{k}A^{0}}%
+(c^{k}-c^{j})\frac{\partial _{jk}A^{0}}{\partial _{j}A^{0}\partial _{k}A^{0}%
}+(c^{j}-c^{i})\frac{\partial _{ij}A^{0}}{\partial _{i}A^{0}\partial
_{j}A^{0}}=0\text{, \ \ }i\neq j\neq k.
\end{equation*}%
However, this is not an unique conservative form. For instance, the
hydrodynamic type system (\textbf{\ref{reda}}) can be written in the form
(see (\textbf{\ref{bm}}) and (\textbf{\ref{first}}))%
\begin{eqnarray*}
A_{t}^{k} &=&A_{x}^{k+1}+kA^{k-1}A_{x}^{0}\text{, \ \ }k=0,1,...,K-2\text{,
\ \ \ \ \ }A_{t}^{K-1}=\partial _{x}A^{K}(\mathbf{c})+(K-1)A^{K-2}A_{x}^{0}%
\text{,} \\
&& \\
c_{t}^{i} &=&\partial _{x}\left( \frac{(c^{i})^{2}}{2}+A^{0}\right) \text{,\
\ \ }i=1,2,...,M,
\end{eqnarray*}%
where $A^{K}(\mathbf{c})$ can be found from the consistency of the above
hydrodynamic type system and its generating function of conservation laws (%
\textbf{\ref{zaks}}) (cf. (\textbf{\ref{egt}}), where $K=0$).

\textbf{Main theorem \cite{Maks+flat}}: \textit{the most general \textbf{%
Hamiltonian} hydrodynamic reduction} (\textbf{\ref{hama}})%
\begin{equation}
c_{t}^{i}=\frac{1}{\varepsilon _{i}}\partial _{x}\frac{\partial \mathbf{h}%
_{K+2}}{\partial c^{i}}\text{, \ \ \ \ \ \ \ \ }\partial _{t}h_{k}=\partial
_{x}\frac{\partial \mathbf{h}_{K+2}}{\partial h_{K-1-k}}\text{, \ \ \ }%
k=0,1,2,...,K-1  \label{ha}
\end{equation}%
\textit{is connected with the equation of the Riemann surface}%
\begin{equation}
\lambda =\frac{\mu ^{K+1}}{K+1}+\underset{k=0}{\overset{K-1}{\sum }}%
Q_{K-1-k}(\mathbf{A})\mu ^{k}-\underset{m=1}{\overset{M}{\sum }}\varepsilon
_{m}\ln (\mu -c^{m}),  \label{ful}
\end{equation}%
\textit{where} $\varepsilon _{k}$ \textit{are arbitrary constants, }$h_{k}(%
\mathbf{A})$ \textit{and} $Q_{k}(\mathbf{A})$ \textit{are some polynomials
with respect to the moments} $A^{n}$.

If $\Sigma \varepsilon _{k}=0$, then $Q_{k}(\mathbf{A})$ can be obtained by
a comparison of the above expression (\textbf{\ref{hama}}) with (\textbf{\ref%
{rm}})%
\begin{equation*}
\frac{\mu ^{K+1}}{K+1}+\underset{k=0}{\overset{K-1}{\sum }}Q_{K-1-k}(\mathbf{%
A})\mu ^{k}-\underset{m=1}{\overset{M}{\sum }}\varepsilon _{m}\ln (\mu
-c^{m})=\frac{1}{K+1}\left( \mu +\underset{k=0}{\overset{\infty }{\sum }}%
\frac{A^{k}}{\mu ^{k+1}}\right) ^{K+1}.
\end{equation*}%
Then all higher moments $A^{k}$ are given by the moment decomposition%
\begin{equation*}
A^{k}=\frac{1}{k+1}\underset{m=1}{\overset{M}{\sum }}\varepsilon
_{m}(c^{m})^{k+1}\text{, \ \ \ \ \ \ \ \ }k=N,N+1,...
\end{equation*}%
If $\Sigma \varepsilon _{k}\neq 0$, then at first the parameter $\lambda $
must be replaced by the combination $\lambda (\mu )\rightarrow \lambda (\mu
)-\Sigma \varepsilon _{k}\ln \lambda (\mu )$.

The Hamiltonian hydrodynamic type system (\textbf{\ref{ha}}) (was found by
L.V. Bogdanov and B.G. Konopelchenko in \textbf{\cite{Bogdan}}) has two 
\textit{most degenerate} limits: All constants $\varepsilon _{k}=0$. Thus,
this is nothing but the Lax hydrodynamic reduction (see \textbf{\cite{Dubr}}
and \textbf{\cite{Krich}}, the dispersionless limit of the Gelfand--Dikey
linear problem). This hydrodynamic reduction can be obtained by the sole
constraint $\mathbf{h}_{K}=0$, then $\mathbf{h}_{K+1}=\Sigma
h_{k}h_{K-1-k}/2 $ (see (\textbf{\ref{2}})) is the momentum density.

The second case is the Kodama hydrodynamic reduction (see \textbf{\cite%
{Kodama}}) associated with the equation of the Riemann surface%
\begin{equation}
\lambda =\frac{\mu ^{K+1}}{K+1}+\underset{k=0}{\overset{K-1}{\sum }}%
Q_{K-1-k}(\mathbf{A})\mu ^{k}+\underset{n=1}{\overset{L}{\sum }}\frac{b_{n}}{%
(\mu -c)^{n}},  \label{kodama}
\end{equation}%
which is sub-case of the Krichever hydrodynamic reduction (see \textbf{\cite%
{Krich}}, and also \textbf{\cite{Aoyama}}) associated with the equation of
the Riemann surface%
\begin{equation}
\lambda =\frac{\mu ^{K+1}}{K+1}+\underset{k=0}{\overset{K-1}{\sum }}%
Q_{K-1-k}(\mathbf{A})\mu ^{k}+\underset{m=1}{\overset{M}{\sum }}\underset{n=1%
}{\overset{N_{m}}{\sum }}\frac{b_{m,n}}{(\mu -c^{m})^{n}}.  \label{krich}
\end{equation}

\section{The waterbag reduction and the WDVV equation}

Let us consider the hydrodynamic reduction (\textbf{\ref{ha}}), where $K=0$
and $M$ is arbitrary.

This is the so-called waterbag hydrodynamic reduction (see \textbf{\cite%
{Gib+Yu}}, \textbf{\cite{Kodama}})%
\begin{equation}
c_{t}^{i}=\partial _{x}\left( \frac{(c^{i})^{2}}{2}+\sum \varepsilon
_{k}c^{k}\right) .  \label{water}
\end{equation}%
In such a case the equation of the Riemann surface is given by (\textbf{\ref%
{ful}})%
\begin{equation}
\lambda =\mu -\underset{m=1}{\overset{M}{\sum }}\varepsilon _{m}\ln (\mu
-c^{m}).  \label{log}
\end{equation}%
In this paper for simplicity we restrict our consideration on the case $%
\Sigma \varepsilon _{k}=0$ (if $\Sigma \varepsilon _{k}\neq 0$, then
corresponding computation became more complicated, but the approach
presented below remains valid).

\textbf{1}. Since $\lambda (x,t)$ is a solution of the linear equation (%
\textbf{\ref{5}}), then $\lambda $ can be replaced by any function $\tilde{%
\lambda}(\lambda )$. Thus, (\textbf{\ref{log}}) can be written in the form%
\begin{equation*}
\lambda =(\mu -c^{i})e^{-\mu /\varepsilon _{i}}\underset{k\neq i}{\prod }%
(\mu -c^{k})^{\varepsilon _{k}/\varepsilon _{i}}
\end{equation*}%
for any fixed index $i$. Then $N$ infinite series of conservation laws%
\begin{equation}
\mu ^{(i)}=c^{i}+h_{i}^{(1)}(\mathbf{c})\lambda +h_{i}^{(2)}(\mathbf{c}%
)\lambda ^{2}+h_{i}^{(3)}(\mathbf{c})\lambda ^{3}+...  \label{bls}
\end{equation}%
can be obtained with the aid of the B\"{u}rmann--Lagrange series (see 
\textbf{\cite{Lavr}}), whose coefficients are determined by%
\begin{equation*}
h_{i}^{(n)}=\frac{1}{n!}\frac{d^{n-1}}{d(c^{i})^{n-1}}\left(
e^{nc^{i}/\varepsilon _{i}}\underset{k\neq i}{\prod }(c^{i}-c^{k})^{-n%
\varepsilon _{k}/\varepsilon _{i}}\right) \text{, \ \ \ \ \ \ \ }n=1,2,...
\end{equation*}%
For instance, the first conservation law densities are%
\begin{equation*}
h_{i}^{(1)}=e^{c^{i}/\varepsilon _{i}}\underset{k\neq i}{\prod }%
(c^{i}-c^{k})^{-\varepsilon _{k}/\varepsilon _{i}}\text{, \ \ \ }h_{i}^{(2)}=%
\frac{e^{2c^{i}/\varepsilon _{i}}}{\varepsilon _{i}}\left( 1-\underset{n\neq
i}{\sum }\frac{\varepsilon _{n}}{c^{i}-c^{n}}\right) \underset{k\neq i}{%
\prod }(c^{i}-c^{k})^{-2\varepsilon _{k}/\varepsilon _{i}},...
\end{equation*}

All above conservation law densities are found by a differentiation only.
Below we develop the approach successfully applied to the Zakharov
hydrodynamic reductions of the Benney hydrodynamic chain (see \textbf{\cite%
{Maks+Benney}}) for construction of conservation law densities for the
Egorov Hamiltonian hydrodynamic type systems.

\textbf{2}. The generating function of conservation laws (\textbf{\ref{zaks}}%
) is consistent with any hydrodynamic reduction written via the Riemann
coordinates (see (\textbf{\ref{rims}}))%
\begin{equation}
r_{t}^{i}=\mu ^{i}(\mathbf{r})r_{x}^{i}\text{, \ \ \ \ \ \ \ }i=1,2,...,N.
\label{dik}
\end{equation}%
It means that the generating function of conservation law densities $\mu $
satisfies the nonlinear PDE system of the first order%
\begin{equation}
\partial _{i}\mu =\frac{\partial _{i}A^{0}}{\mu ^{i}-\mu }.  \label{geni}
\end{equation}%
At the same time, the moment $A^{0}(\mathbf{r})$ is a potential of the
Egorov metric (see (\textbf{\ref{ega}}) and \textbf{\cite{Maks+Egor+int}}).
Then an arbitrary commuting flow%
\begin{equation}
r_{\tau }^{i}=w^{i}(\mathbf{r})r_{x}^{i}\text{, \ \ \ \ \ \ \ }i=1,2,...,N
\label{coma}
\end{equation}%
has the corresponding conservation law $A_{\tau }^{0}=h_{x}$, where $h(%
\mathbf{r})$ is a conservation law density.

\textbf{Lemma}: \textit{The generating function of commuting flows} (\textbf{%
\ref{coma}})%
\begin{equation}
r_{\tau (\zeta )}^{i}=\frac{1}{\mu ^{i}-\mu }r_{x}^{i}\text{, \ \ \ \ \ \ \ }%
i=1,2,...,N  \label{cona}
\end{equation}%
\textit{is connected with the conservation law} $\partial _{\tau (\zeta
)}A^{0}=\mu _{x}$. \textit{The generating function of conservation laws and
commuting flows is given by}%
\begin{equation}
\partial _{\tau (\zeta )}\mu (\lambda )=\partial _{x}\ln [\mu (\lambda )-\mu
(\zeta )].  \label{son}
\end{equation}

\textbf{Proof}: Let us differentiate the conservation law $A_{\tau (\zeta
)}^{0}=\partial _{x}\mu (\zeta )$ with respect to the Riemann invariants.
Then (see (\textbf{\ref{coma}}))%
\begin{equation*}
w^{i}=\frac{\partial _{i}\mu }{\partial _{i}A^{0}},
\end{equation*}%
and (\textbf{\ref{cona}}) can be obtained taking into account (\textbf{\ref%
{geni}}). Differentiation of the generating function (\textbf{\ref{son}})
with respect to the Riemann invariants yields the same result (\textbf{\ref%
{cona}}).

\textbf{3}. $N$ \textit{primary} commuting flows in the Riemann invariants
(see (\textbf{\ref{bls}}))%
\begin{equation}
r_{t^{k}}^{i}=\frac{1}{\mu ^{i}-c^{k}}r_{x}^{i}\text{, \ \ \ \ \ \ \ }%
i=1,2,...,N  \label{rima}
\end{equation}%
can be obtained directly from the generating function (\textbf{\ref{cona}})
according $N$ punctures $c^{i}$, where $\mu (\zeta )\rightarrow c^{k}$ and $%
\tau (\zeta )\rightarrow t^{k}$ (see (\textbf{\ref{log}})). These commuting
flows in the conservative form are%
\begin{equation*}
\mu _{t^{k}}=\partial _{x}\ln (\mu -c^{k}).
\end{equation*}%
Another choice of punctures $c^{i}$, where $\mu (\lambda )\rightarrow c^{i}$
leads to the generating function of commuting flows%
\begin{equation}
c_{\tau }^{i}=\partial _{x}\ln (c^{i}-\mu ),  \label{gema}
\end{equation}%
while the hydrodynamic type system (\textbf{\ref{dik}}) in the conservative
form is given by (\textbf{\ref{water}}).

\textbf{4}. Since the waterbag reduction (\textbf{\ref{water}}) is the
Hamiltonian hydrodynamic type system%
\begin{equation}
c_{t}^{k}=\frac{1}{\varepsilon _{k}}\partial _{x}\frac{\partial \mathbf{h}%
_{2}}{\partial c^{k}},  \label{ham}
\end{equation}%
where $\mathbf{h}_{2}=A^{2}+(A^{0})^{2}$, then each commuting flow has the
same local Hamiltonian structure%
\begin{equation}
c_{t_{p}^{s}}^{k}=\frac{1}{\varepsilon _{k}}\partial _{x}\frac{\partial 
\mathbf{h}_{s}^{(p)}}{\partial c^{k}}\text{, \ \ \ \ }k,s=1,2,...,N\text{, \
\ }p=0,1,2,...  \label{fuki}
\end{equation}

\textbf{Lemma}: \textit{The generating function of commuting flows} (\textbf{%
\ref{gema}}) \textit{is determined by the Hamiltonian density}%
\begin{equation}
\mathbf{h}(\zeta )=\frac{\mu ^{2}(\zeta )}{2}+A^{0}-\sum \varepsilon
_{k}c^{k}\ln [\mu (\zeta )-c^{k}].  \label{6}
\end{equation}

\textbf{Proof}: Since the generating function of commuting flows (\textbf{%
\ref{gema}}) has the same Hamiltonian structure (\textbf{\ref{ham}}), then
an integration of the differential (see (\textbf{\ref{gema}}) and (\textbf{%
\ref{ham}}))%
\begin{equation*}
d\mathbf{h}(\zeta )=\sum \varepsilon _{k}\ln [\mu (\zeta )-c^{k}]dc^{k}
\end{equation*}%
yields (\textbf{\ref{6}}).

The term $A^{0}$ in the Hamiltonian density (\textbf{\ref{6}}) is
inessential. By this reason we remove it from a further consideration. At
the beginning of this section we mentioned the existence on $N+2$ infinite
series of conservation laws. The substitution of the Kruskal series in (%
\textbf{\ref{6}}) yields the same set of the Kruskal conservation laws.
However, the substitution of the B\"{u}rmann--Lagrange series (\textbf{\ref%
{bls}}) at the first step yields the logaritmic conservation law densities%
\begin{equation}
\tilde{h}_{i}^{(1)}=\frac{(c^{i})^{2}}{2}+\underset{k\neq i}{\sum }%
\varepsilon _{k}(c^{k}-c^{i})\ln (c^{k}-c^{i}).  \label{ln}
\end{equation}%
The corresponding hydrodynamic type systems are (\textbf{\ref{rima}}).

\textbf{5}. $A^{0}=\Sigma \varepsilon _{k}c^{k}$ is a potential of the
Egorov metric (see (\textbf{\ref{ega}}), (\textbf{\ref{ham}}) and (\textbf{%
\ref{fuki}}))%
\begin{equation*}
A_{t}^{0}=\partial _{x}(\delta \mathbf{h}_{2})=A_{x}^{1}\text{, \ \ \ \ \ }%
A_{t_{p}^{s}}^{0}=\partial _{x}(\delta \tilde{h}_{s}^{(p)})=\partial _{x}%
\tilde{h}_{s}^{(p-1)},
\end{equation*}%
where $\delta =\Sigma \partial /\partial c^{k}$ is a shift operator and $%
\tilde{h}_{i}^{(0)}=c^{i}$. Since $A_{\tau (\zeta )}^{0}=\mu _{x}$, then it
means that $\delta \mathbf{h}_{p+1}=\mathbf{h}_{p}$, $\delta
h_{s}^{(p)}=h_{s}^{(p-1)}$, $p=0,1,2,...$, $s=1,2,...,N$.

\textbf{Lemma}: $N$ \textit{commuting flows} (\textbf{\ref{thrid}})%
\begin{equation}
c_{t^{k}}^{i}=\frac{1}{\varepsilon _{i}}\partial _{x}\frac{\partial \tilde{h}%
_{k}^{(1)}}{\partial c^{i}}  \label{stop}
\end{equation}%
\textit{are determined by the conservation laws}%
\begin{equation*}
A_{t^{k}}^{0}=\partial _{x}c^{k},
\end{equation*}%
\textit{where} $t_{1}^{k}\equiv t^{k}$.

\textbf{Proof}: Indeed, $\delta \tilde{h}_{k}^{(1)}=\tilde{h}%
_{k}^{(0)}=c^{k} $ (see (\textbf{\ref{ln}})).

Let us write the above commuting flows (\textbf{\ref{stop}}) in the
potential form (\textbf{\ref{dif}})%
\begin{equation*}
d\xi ^{i}=c^{i}dx+\underset{k\neq i}{\sum }\ln (c^{i}-c^{k})dt^{k}+\frac{1}{%
\varepsilon _{i}}\left( c^{i}-\underset{k\neq i}{\sum }\ln
(c^{i}-c^{k})\right) dt^{i}.
\end{equation*}%
Let us introduce new flat coordinates (\textbf{\ref{flet}})%
\begin{equation*}
a_{1}=\frac{1}{\varepsilon _{1}}\left( c^{1}-\underset{k\neq 1}{\sum }%
\varepsilon _{k}\ln (c^{1}-c^{k})\right) \text{, \ \ \ \ \ \ \ }a_{k}=\ln
(c^{1}-c^{k})\text{, \ \ }k=2,3,...,N.
\end{equation*}%
Since, the above point transformation is invertible%
\begin{equation*}
c^{1}=\sum \varepsilon _{n}a_{n}\text{, \ \ \ \ \ \ \ \ }c^{k}=\sum
\varepsilon _{n}a_{n}-e^{a_{k}}\text{, \ \ }k=2,3,...,N
\end{equation*}%
the canonical Egorov basic set is given by%
\begin{eqnarray}
\partial _{t^{n}}a_{1} &=&\partial _{t^{1}}a_{n}\text{, \ \ \ \ \ \ \ }%
\partial _{t^{n}}a_{k}=\partial _{t^{1}}\ln (e^{a_{k}}-e^{a_{n}})\text{, \ \
\ \ }k\neq 1,n,  \notag \\
&&  \label{l} \\
\partial _{t^{n}}a_{n} &=&\frac{1}{\varepsilon _{n}}\partial _{t^{1}}\left(
\sum \varepsilon _{m}a_{m}-\varepsilon _{1}a_{n}-e^{a_{n}}-\underset{m\neq
1,n}{\sum }\varepsilon _{m}\ln (e^{a_{m}}-e^{a_{n}})\right) .  \notag
\end{eqnarray}%
Thus, the above hydrodynamic type systems can be written in the form (%
\textbf{\ref{zima}}), where $a^{k}=\varepsilon _{k}a_{k}$. Then the
corresponding solution of the WDVV equation (\textbf{\ref{wdvv}})%
\begin{equation}
\sum \frac{1}{\varepsilon _{s}}\left( \frac{\partial ^{3}F}{\partial
a_{k}\partial a_{i}\partial a_{s}}\frac{\partial ^{3}F}{\partial
a_{s}\partial a_{j}\partial a_{n}}-\frac{\partial ^{3}F}{\partial
a_{j}\partial a_{i}\partial a_{s}}\frac{\partial ^{3}F}{\partial
a_{s}\partial a_{k}\partial a_{n}}\right) =0  \label{new}
\end{equation}%
can be found in quadratures%
\begin{equation*}
F=\frac{\varepsilon _{1}^{2}(a_{1})^{3}}{6}+\frac{\varepsilon _{1}a_{1}}{2}%
\sum \varepsilon _{m}(a_{m})^{2}+P_{3}(\mathbf{a})-\sum \varepsilon
_{m}e^{a_{m}}+\frac{1}{2}\underset{m<k}{\sum }\varepsilon _{k}\varepsilon
_{m}\left[ \text{Li}_{3}\left( e^{a_{k}-a_{m}}\right) +\text{Li}_{3}\left(
e^{a_{m}-a_{k}}\right) \right] ,
\end{equation*}%
where%
\begin{equation*}
P_{3}(\mathbf{a})=\sum \frac{\varepsilon _{m}(\varepsilon _{m}-\varepsilon
_{1})(a_{m})^{3}}{6}+\underset{m<k}{\sum }\frac{\varepsilon _{k}\varepsilon
_{m}}{12}\left[ (a_{k}+a_{m})^{3}-2((a_{k})^{3}+(a_{m})^{3})\right] .
\end{equation*}%
\textbf{Attention}: the index $1$ \textbf{is not included} in the above two
summations.

For instance, if $N=3$, then%
\begin{eqnarray}
F &=&\frac{\varepsilon _{1}^{2}(a_{1})^{3}}{6}+\frac{\varepsilon _{1}a_{1}}{2%
}[\varepsilon _{2}(a_{2})^{2}+\varepsilon _{3}(a_{3})^{2}]+\frac{\varepsilon
_{2}\varepsilon _{3}}{4}[a_{3}(a_{2})^{2}+a_{2}(a_{3})^{2}]  \notag \\
&&  \notag \\
&&+\left( \frac{\varepsilon _{2}(\varepsilon _{2}-\varepsilon _{1})}{6}-%
\frac{\varepsilon _{2}\varepsilon _{3}}{12}\right) (a_{2})^{3}+\left( \frac{%
\varepsilon _{3}(\varepsilon _{3}-\varepsilon _{1})}{6}-\frac{\varepsilon
_{2}\varepsilon _{3}}{12}\right) (a_{3})^{3}  \label{j} \\
&&  \notag \\
&&-\varepsilon _{2}e^{a_{2}}-\varepsilon _{3}e^{a_{3}}+\frac{\varepsilon
_{2}\varepsilon _{3}}{2}\left[ \text{Li}_{3}\left( e^{a_{3}-a_{2}}\right) +%
\text{Li}_{3}\left( e^{a_{2}-a_{3}}\right) \right] .  \notag
\end{eqnarray}

\textbf{Remark}: Similar (but different!) solutions are found by R. Martini
and L.K. Hoevenaars in \textbf{\cite{Martini}}.

\section{Degenerations of the waterbag hydrodynamic reduction}

The waterbag hydrodynamic reduction is a most general Hamiltonian
hydrodynamic reduction (\textbf{\ref{hama}}), which can be obtained by the 
\textit{Dirac restriction} (see \textbf{\cite{Fer+trans}}) of the
Kupershmidt--Manin Poisson bracket for the Benney hydrodynamic chain (see 
\textbf{\cite{Maks+flat}}). All other ``flat'' hydrodynamic reductions
connected with the equation of the Riemann surface%
\begin{equation*}
\lambda =\mu -\underset{k=1}{\overset{N_{0}}{\sum }}\varepsilon _{k}\ln (\mu
-a_{0}^{k})+\underset{m=1}{\overset{M}{\sum }}\underset{k=1}{\overset{N_{m}}{%
\sum }}\frac{\tilde{a}_{m}^{k}}{(\mu -a_{m}^{k})^{m}}
\end{equation*}%
and obtained by the Dirac restriction of the Kupershmidt--Manin Poisson
bracket can be derived from the waterbag hydrodynamic reduction by an
appropriate degeneration (see \textbf{\cite{Maks+flat}}). In this case $M$
is an arbitrary integer (the number of equations in corresponding
hydrodynamic reduction is $N=\underset{m=0}{\overset{M}{\sum }}(m+1)N_{m}$,
where $N_{m}$ are arbitrary integers). For instance, for $M=3$ we have%
\begin{eqnarray*}
\lambda &=&\mu -\underset{k=1}{\overset{N_{0}}{\sum }}\varepsilon _{k}\ln
(\mu -a_{0}^{k})+\underset{k=1}{\overset{N_{1}}{\sum }}\frac{c_{1}^{k}}{\mu
-a_{1}^{k}}+\underset{k=1}{\overset{N_{2}}{\sum }}\frac{c_{2}^{k}}{\mu
-a_{2}^{k}}+\frac{1}{2}\underset{k=1}{\overset{N_{2}}{\sum }}\frac{%
(b_{2}^{k})^{2}}{(\mu -a_{2}^{k})^{2}}+ \\
&& \\
&&+\underset{k=1}{\overset{N_{3}}{\sum }}\frac{d_{3}^{k}}{\mu -a_{3}^{k}}+%
\underset{k=1}{\overset{N_{3}}{\sum }}\frac{c_{3}^{k}b_{3}^{k}}{(\mu
-a_{3}^{k})^{2}}+\frac{1}{3}\underset{k=1}{\overset{N_{3}}{\sum }}\frac{%
(b_{3}^{k})^{3}}{(\mu -a_{3}^{k})^{3}},
\end{eqnarray*}%
where $N_{0}$, $N_{1}$, $N_{2}$ and $N_{3}$ are arbitrary integers.
Corresponding hydrodynamic type system%
\begin{eqnarray*}
\partial _{t}a_{0}^{k} &=&\partial _{x}[\frac{(a_{0}^{k})^{2}}{2}+A^{0}]%
\text{, \ \ \ \ \ \ }\partial _{t}a_{1}^{k}=\partial _{x}[\frac{%
(a_{1}^{k})^{2}}{2}+A^{0}]\text{, \ \ \ \ \ \ }\partial
_{t}b_{1}^{k}=\partial _{x}(b_{1}^{k}a_{1}^{k})\text{,} \\
&& \\
\partial _{t}a_{2}^{k} &=&\partial _{x}[\frac{(a_{2}^{k})^{2}}{2}+A^{0}]%
\text{, \ \ \ \ \ \ }\partial _{t}b_{2}^{k}=\partial _{x}(b_{2}^{k}a_{2}^{k})%
\text{, \ \ \ \ \ \ }\partial _{t}c_{2}^{k}=\partial _{x}[a_{2}^{k}c_{2}^{k}+%
\frac{1}{2}(b_{2}^{k})^{2}], \\
&& \\
\partial _{t}a_{3}^{k} &=&\partial _{x}[\frac{(a_{3}^{k})^{2}}{2}+A^{0}]%
\text{, \ \ }\partial _{t}b_{3}^{k}=\partial _{x}(b_{3}^{k}a_{3}^{k})\text{,
\ \ }\partial _{t}c_{3}^{k}=\partial _{x}[a_{3}^{k}c_{3}^{k}+\frac{1}{2}%
(b_{3}^{k})^{2}]\text{, \ }\partial _{t}d_{3}^{k}=\partial
_{x}[a_{3}^{k}d_{3}^{k}+b_{3}^{k}c_{3}^{k}],
\end{eqnarray*}%
where $A^{0}=\underset{k=1}{\overset{N_{0}}{\sum }}\varepsilon _{k}a_{0}^{k}+%
\underset{k=1}{\overset{N_{1}}{\sum }}b_{1}^{k}+\underset{k=1}{\overset{N_{2}%
}{\sum }}c_{2}^{k}+\underset{k=1}{\overset{N_{3}}{\sum }}d_{3}^{k}$, has
following local Hamiltonian structure%
\begin{eqnarray*}
\partial _{t}a_{0}^{k} &=&\frac{1}{2}\partial _{x}[\frac{1}{\varepsilon _{k}}%
\frac{\partial \mathbf{h}_{2}}{\partial a_{0}^{k}}]\text{, \ \ \ \ \ }%
\partial _{t}a_{1}^{k}=\frac{1}{2}\partial _{x}\frac{\partial \mathbf{h}_{2}%
}{\partial b_{1}^{k}}\text{, \ \ \ \ \ \ }\partial _{t}b_{1}^{k}=\frac{1}{2}%
\partial _{x}\frac{\partial \mathbf{h}_{2}}{\partial a_{1}^{k}}, \\
&& \\
\partial _{t}a_{2}^{k} &=&\frac{1}{2}\partial _{x}\frac{\partial \mathbf{h}%
_{2}}{\partial c_{1}^{k}}\text{, \ \ \ \ \ }\partial _{t}b_{2}^{k}=\frac{1}{2%
}\partial _{x}\frac{\partial \mathbf{h}_{2}}{\partial b_{2}^{k}}\text{, \ \
\ \ \ \ }\partial _{t}c_{2}^{k}=\frac{1}{2}\partial _{x}\frac{\partial 
\mathbf{h}_{2}}{\partial a_{2}^{k}}, \\
&& \\
\partial _{t}a_{3}^{k} &=&\frac{1}{2}\partial _{x}\frac{\partial \mathbf{h}%
_{2}}{\partial d_{3}^{k}}\text{, \ \ \ \ \ }\partial _{t}b_{3}^{k}=\frac{1}{2%
}\partial _{x}\frac{\partial \mathbf{h}_{2}}{\partial c_{3}^{k}}\text{, \ \
\ \ \ \ }\partial _{t}c_{3}^{k}=\frac{1}{2}\partial _{x}\frac{\partial 
\mathbf{h}_{2}}{\partial b_{3}^{k}}\text{, \ \ \ \ \ \ }\partial
_{t}d_{3}^{k}=\frac{1}{2}\partial _{x}\frac{\partial \mathbf{h}_{2}}{%
\partial a_{3}^{k}},
\end{eqnarray*}%
where%
\begin{eqnarray*}
A^{n}(\mathbf{a},\mathbf{b},\mathbf{c},\mathbf{d}) &=&\frac{1}{n+1}\underset{%
k=1}{\overset{N_{0}}{\sum }}\varepsilon _{k}(a_{0}^{k})^{n+1}+\underset{k=1}{%
\overset{N_{1}}{\sum }}(a_{1}^{k})^{n}b_{1}^{k}+\underset{k=1}{\overset{N_{2}%
}{\sum }}(a_{2}^{k})^{n}c_{2}^{k}+\frac{n}{2}\underset{k=1}{\overset{N_{2}}{%
\sum }}(a_{2}^{k})^{n-1}(b_{2}^{k})^{2} \\
&& \\
&&+\underset{k=1}{\overset{N_{3}}{\sum }}(a_{3}^{k})^{n}d_{3}^{k}+n\underset{%
k=1}{\overset{N_{3}}{\sum }}(a_{3}^{k})^{n-1}b_{3}^{k}c_{3}^{k}+\frac{n(n-1)%
}{6}\underset{k=1}{\overset{N_{3}}{\sum }}(a_{3}^{k})^{n-2}(b_{3}^{k})^{3}.
\end{eqnarray*}

The Benney hydrodynamic chain has an infinite series of local Hamiltonian
structures (see \textbf{\cite{Maks+flat}}). The most general Hamiltonian
hydrodynamic reductions of the Benney hydrodynamic chain obtained by the
Dirac restriction of $(K+1)$st local Hamiltonian structure are determined by
the equation of the Riemann surface (\textbf{\ref{ful}}). Its degenerations
are given by the more complicated equation%
\begin{equation*}
\lambda =\frac{\mu ^{K+1}}{K+1}+\underset{k=0}{\overset{K-1}{\sum }}%
A_{(K)}^{k}\mu ^{K-(k+1)}-\underset{k=1}{\overset{N_{0}}{\sum }}\varepsilon
_{k}\ln (\mu -a_{0}^{k})+\underset{m=1}{\overset{M}{\sum }}\underset{k=1}{%
\overset{N_{m}}{\sum }}\frac{\tilde{a}_{m}^{k}}{(\mu -a_{m}^{k})^{m}}.
\end{equation*}%
In this case the number of equations in the corresponding hydrodynamic type
system is $N=K+\underset{m=0}{\overset{M}{\sum }}(m+1)N_{m}$, where $N_{m}$
are arbitrary integers. This above equation was found by L.V. Bogdanov and
B.G. Konopelchenko (see \textbf{\cite{Bogdan}}); if all parameters $%
\varepsilon _{k}=0$, this case was considered by I.M. Krichever (see \textbf{%
\cite{Krich}}).

\subsection{The Zakharov hydrodynamic reduction}

The Zakharov hydrodynamic reduction (this is $2N$ component hydrodynamic
type system, see \textbf{\cite{Zakh}})%
\begin{equation*}
u_{t}^{i}=\partial _{x}\left( \frac{(u^{i})^{2}}{2}+\sum \eta ^{m}\right) 
\text{, \ \ \ \ \ \ }\eta _{t}^{i}=\partial _{x}(u^{i}\eta ^{i})\text{, \ \
\ }i=1,2,...,N
\end{equation*}%
is associated with the equation of the Riemann surface%
\begin{equation*}
\lambda =\mu +\sum \frac{\eta ^{n}}{\mu -u^{n}}.
\end{equation*}%
A substitution of the inverse series (at the vicinity of the puncture $\mu
^{(k)}=u^{k}$)%
\begin{equation*}
\mu ^{(k)}=u^{k}+\eta ^{k}/\lambda +h_{k}^{(2)}(\mathbf{u,\eta })/\lambda
^{2}+h_{k}^{(3)}(\mathbf{u,\eta })/\lambda ^{3}+...
\end{equation*}%
in the above equation yields \textit{first} $N$ series of conservation law
densities, where the \textit{first} $N$ conservation law densities are%
\begin{equation*}
h_{k}^{(2)}=\eta ^{k}\left( u^{k}+\underset{m\neq k}{\sum }\frac{\eta ^{m}}{%
u^{k}-u^{m}}\right) .
\end{equation*}%
A substitution of the above inverse series in (\textbf{\ref{cona}}) yields $%
2N$ primary commuting flows written in the Riemann invariants (cf. (\textbf{%
\ref{rima}}))%
\begin{equation}
r_{t^{k}}^{i}=\frac{1}{\mu ^{i}-u^{k}}r_{x}^{i}\text{, \ \ \ \ \ \ }%
r_{t^{k+N}}^{i}=\frac{\eta ^{k}}{(\mu ^{i}-u^{k})^{2}}r_{x}^{i}\text{,\ \ \
\ \ }i=1,2,...,N.  \label{min}
\end{equation}%
A substitution of the above inverse series in (\textbf{\ref{son}}) yields a
generating function of conservation laws for these $2N$ commuting flows
written in the conservative form%
\begin{equation}
\mu _{t^{k}}=\partial _{x}\ln (\mu -u^{k}),\ \ \ \ \ \ \ \mu
_{t^{k+N}}=\partial _{x}\frac{\eta ^{k}}{u^{k}-\mu }.  \label{prim}
\end{equation}%
A substitution of the above inverse series in (\textbf{\ref{son}}) yields $%
2N $ generating functions of commuting flows in the conservative form%
\begin{equation*}
u_{\tau }^{i}=\partial _{x}\ln (u^{i}-\mu )\text{, \ \ \ \ \ \ \ }\eta
_{\tau }^{i}=\partial _{x}\frac{\eta ^{i}}{u^{i}-\mu }.
\end{equation*}%
Since the Zakharov hydrodynamic reduction has the local Hamiltonian structure%
\begin{equation*}
u_{t}^{i}=\partial _{x}\frac{\partial \mathbf{h}_{2}}{\partial \eta ^{i}}%
\text{, \ \ \ \ \ \ \ }\eta _{t}^{i}=\partial _{x}\frac{\partial \mathbf{h}%
_{2}}{\partial u^{i}},
\end{equation*}%
then the above generating functions of commuting flows are determined by the
same Hamiltonian structure, where the Hamiltonian density is given by (cf. (%
\textbf{\ref{6}}))%
\begin{equation*}
\mathbf{h}(\lambda )=\frac{\mu ^{2}(\lambda )}{2}+\sum \eta ^{k}\ln
[u^{k}-\mu (\lambda )].
\end{equation*}%
A substitution of the above inverse series in this generating function of
conservation law densities yields \textit{second} $N$ series of conservation
law densities, where the \textit{second} $N$ conservation law densities are
(see \textbf{\cite{Maks+Benney}})%
\begin{equation*}
h_{k+N}^{(2)}=\frac{(u^{k})^{2}}{2}+\eta ^{k}\ln \eta ^{k}+\underset{m\neq k}%
{\sum }\eta ^{m}\ln (u^{k}-u^{m}).
\end{equation*}%
Then $2N$ primary commuting flows (\textbf{\ref{min}}) in the conservative
form%
\begin{eqnarray*}
u_{t^{k}}^{i} &=&\partial _{x}\frac{\eta ^{k}}{u^{k}-u^{i}}\text{,\ \ \ \ \
\ \ \ }\eta _{t^{k}}^{i}=\partial _{x}\frac{\eta ^{i}\eta ^{k}}{%
(u^{k}-u^{i})^{2}}\text{, \ \ \ \ }i\neq k, \\
&& \\
u_{t^{k}}^{k} &=&\partial _{x}\left( u^{k}+\underset{m\neq k}{\sum }\frac{%
\eta ^{m}}{u^{k}-u^{m}}\right) \text{,\ \ \ \ \ \ }\eta
_{t^{k}}^{k}=\partial _{x}\left[ \eta ^{k}\left( 1-\underset{m\neq k}{\sum }%
\frac{\eta ^{m}}{(u^{k}-u^{m})^{2}}\right) \right] \text{,} \\
&& \\
u_{t^{k+N}}^{i} &=&\partial _{x}\ln (u^{k}-u^{i})\text{, \ \ \ \ \ \ }\eta
_{t^{k+N}}^{i}=-\partial _{x}\frac{\eta ^{i}}{u^{k}-u^{i}}\text{, \ \ \ }%
i\neq k, \\
&& \\
u_{t^{k+N}}^{k} &=&\partial _{x}\ln \eta ^{k}\text{, \ \ \ \ \ \ }\eta
_{t^{k+N}}^{k}=\partial _{x}\left( u^{k}+\underset{m\neq k}{\sum }\frac{\eta
^{m}}{u^{k}-u^{m}}\right)
\end{eqnarray*}%
are determined by the Hamiltonian densities%
\begin{equation*}
u_{t^{k}}^{i}=\partial _{x}\frac{\partial \mathbf{h}_{k}^{(2)}}{\partial
\eta ^{i}}\text{,\ \ \ \ }\eta _{t^{k}}^{i}=\partial _{x}\frac{\partial 
\mathbf{h}_{k}^{(2)}}{\partial u^{i}}\text{,\ \ \ \ }u_{t^{k+N}}^{i}=%
\partial _{x}\frac{\partial \mathbf{h}_{k+N}^{(2)}}{\partial \eta ^{i}}\text{%
,\ \ \ \ }\eta _{t^{k}}^{i}=\partial _{x}\frac{\partial \mathbf{h}%
_{k+N}^{(2)}}{\partial u^{i}}\text{, \ \ }i,k=1,2,...,N.
\end{equation*}%
The first $2N$ primary commuting flows are determined by the extra
conservation law (see (\textbf{\ref{ek}}), cf. (\textbf{\ref{stop}}))%
\begin{equation}
A_{t^{k}}^{0}=\eta _{x}^{k};  \label{raz}
\end{equation}%
the second $2N$ primary commuting flows are determined by the extra
conservation law (see (\textbf{\ref{ek}}), cf. (\textbf{\ref{stop}}) and the
above case)%
\begin{equation}
A_{t^{k+N}}^{0}=u_{x}^{k},  \label{dva}
\end{equation}%
where the potential of the Egorov metric $A^{0}=\Sigma \eta ^{n}$.

\textbf{Remark}: These $2N$ primary commuting flows written in the
conservative form also can be obtained by a direct substitution the above
inverse series in (\textbf{\ref{prim}}).

Let us write these commuting flows in the potential form%
\begin{equation*}
d\xi ^{i}=\eta ^{i}\left( 1-\underset{m\neq i}{\sum }\frac{\eta ^{m}}{%
(u^{i}-u^{m})^{2}}\right) dt^{i}+\left( u^{i}+\underset{m\neq i}{\sum }\frac{%
\eta ^{m}}{u^{i}-u^{m}}\right) dt^{i+N}+\underset{m\neq i}{\sum }\left( 
\frac{\eta ^{i}\eta ^{m}dt^{m}}{(u^{m}-u^{i})^{2}}-\frac{\eta ^{i}dt^{m+N}}{%
u^{m}-u^{i}}\right) ,
\end{equation*}%
\begin{equation*}
d\xi ^{i+N}=\left( u^{i}+\underset{m\neq i}{\sum }\frac{\eta ^{m}}{%
u^{i}-u^{m}}\right) dt^{i}+\ln \eta ^{i}dt^{i+N}+\underset{m\neq i}{\sum }%
\frac{\eta ^{m}}{u^{m}-u^{i}}dt^{m}+\underset{m\neq i}{\sum }\ln
(u^{m}-u^{i})dt^{m+N}\text{.}
\end{equation*}%
In fact we have just two options for derivation of a corresponding solution
for the WDVV equation, because the above differentials are symmetric under
any the permutability $u^{k}\leftrightarrow u^{n}$ and $\eta
^{k}\leftrightarrow \eta ^{n}$.

\textbf{1}. Taking into account (see (\textbf{\ref{sym}}), (\textbf{\ref{raz}%
}) and (\textbf{\ref{dva}}))%
\begin{eqnarray*}
\partial _{j}A^{0} &=&H_{j}^{2}\text{, \ \ \ \ }\partial
_{j}u^{i}=H_{j}^{(i)}H_{j}\text{, \ \ \ \ }\partial _{j}\eta ^{i}=\tilde{H}%
_{j}^{(i)}H_{j}\text{, \ \ \ \ }\partial _{j}\frac{\eta ^{k}}{u^{k}-u^{i}}%
=H_{j}^{(i)}\tilde{H}_{j}^{(k)}, \\
&& \\
\partial _{j}\ln (u^{i}-u^{k}) &=&H_{j}^{(i)}H_{j}^{(k)}\text{, \ \ \ }%
\partial _{j}\frac{\eta ^{i}\eta ^{k}}{(u^{i}-u^{k})^{2}}=\tilde{H}_{j}^{(i)}%
\tilde{H}_{j}^{(k)}\text{, \ \ }\partial _{j}\left( u^{i}+\underset{m\neq i}{%
\sum }\frac{\eta ^{m}}{u^{i}-u^{m}}\right) =H_{j}^{(i)}\tilde{H}_{j}^{(i)},
\\
&& \\
\partial _{j}\ln \eta ^{i} &=&\left( H_{j}^{(i)}\right) ^{2}\text{, \ \ \ \
\ \ \ \ \ \ \ }\partial _{j}\left[ \eta ^{i}\left( 1-\underset{m\neq i}{\sum 
}\frac{\eta ^{m}}{(u^{i}-u^{m})^{2}}\right) \right] =\left( \tilde{H}%
_{j}^{(i)}\right) ^{2}
\end{eqnarray*}%
let us choose new flat coordinates%
\begin{equation*}
a_{1}=\eta ^{1}\left( 1-\underset{m\neq 1}{\sum }\frac{\eta ^{m}}{%
(u^{1}-u^{m})^{2}}\right) \text{, \ \ }a_{k}=\frac{\eta ^{1}\eta ^{k}}{%
(u^{1}-u^{k})^{2}}\text{, \ }b_{1}=u^{1}+\underset{m\neq 1}{\sum }\frac{\eta
^{m}}{u^{1}-u^{m}}\text{, \ \ }b_{k}=\frac{\eta ^{1}}{u^{1}-u^{k}}.
\end{equation*}%
Let us replace $t^{k}\rightarrow b_{k}$ and $t^{k+N}\rightarrow a_{k}$ in
the above potential forms taking into account (\textbf{\ref{ekviv}}).
Respectively, we must replace $\xi ^{i}\rightarrow \partial F/\partial b_{i}$
and $\xi ^{i+N}\rightarrow \partial F/\partial a_{i}$. Then the
corresponding solution of the WDVV equation is given by%
\begin{eqnarray*}
F &=&\frac{a_{1}(b_{1})^{2}}{2}+\frac{\left( a_{1}+\underset{m\neq 1}{\sum }%
a_{m}\right) }{2}\left[ \ln \left( a_{1}+\underset{m\neq 1}{\sum }%
a_{m}\right) -\frac{3}{2}\right] -a_{1}\underset{m\neq 1}{\sum }a_{m}\ln
b_{m} \\
&& \\
&&+b_{1}\underset{m\neq 1}{\sum }a_{m}b_{m}-\frac{1}{2}\underset{m\neq 1}{%
\sum }a_{m}(b_{m})^{2}+\underset{m<k}{\sum }a_{m}a_{k}\ln \frac{a_{m}-a_{k}}{%
b_{m}b_{k}}-\underset{n=1}{\overset{N}{\sum }}(a_{k})^{2}\ln b_{k},
\end{eqnarray*}%
where $a^{k}\equiv b_{k}$ and $b^{k}\equiv a_{k}$.

\textbf{2}. Introducing another set of flat coordinates%
\begin{equation*}
c_{1}=\ln \eta ^{1}\text{, \ \ }c_{k}=\ln (u^{1}-u^{k})\text{, \ \ \ }%
b_{1}=u^{1}+\underset{m\neq 1}{\sum }\frac{\eta ^{m}}{u^{1}-u^{m}}\text{, \
\ \ \ }b_{k}=\frac{\eta ^{k}}{u^{k}-u^{1}}
\end{equation*}%
one can obtain another solution of the WDVV equation connected with the
above solution by the Legendre type transformation (see details in \textbf{%
\cite{Dubr}} and \textbf{\cite{Maks+Egor+int}}).

\subsection{Three component solutions of the WDVV equation}

In this paper without lost of generality we restrict our consideration for
simplicity on three component case. It is easy to enumerate all
corresponding hydrodynamic reductions of the Benney hydrodynamic chain.

\textbf{1}. The waterbag hydrodynamic reduction associated with the equation
of the Riemann surface%
\begin{equation*}
\lambda =\mu -\underset{m=1}{\overset{3}{\sum }}\varepsilon _{k}\ln (\mu
-a^{k})
\end{equation*}%
is considered in the previous section. The corresponding solution (\textbf{%
\ref{j}}) of the WDVV equation is given for $N=3$ also.

\textbf{2}. The first degenerate case 
\begin{equation*}
\lambda =\mu -\varepsilon \ln (\mu -a)+\frac{b}{\mu -c}
\end{equation*}%
can be obtained from the previous waterbag hydrodynamic reduction by a
merging of two singular points. The corresponding Egorov hydrodynamic type
system%
\begin{equation*}
a_{t}=\partial _{x}\left( \frac{a^{2}}{2}+b+\varepsilon a\right) \text{, \ \
\ \ \ \ }b_{t}=\partial _{x}(bc)\text{, \ \ \ \ \ \ }c_{t}=\partial
_{x}\left( \frac{c^{2}}{2}+b+\varepsilon a\right)
\end{equation*}%
possesses the local Hamiltonian structure%
\begin{equation*}
a_{t}=\frac{1}{\varepsilon }\partial _{x}\frac{\partial \mathbf{h}_{2}}{%
\partial a}\text{, \ \ \ \ \ \ \ }b_{t}=\partial _{x}\frac{\partial \mathbf{h%
}_{2}}{\partial c}\text{, \ \ \ \ \ \ \ }c_{t}=\partial _{x}\frac{\partial 
\mathbf{h}_{2}}{\partial b},
\end{equation*}%
and the potential of the Egorov metric is $A^{0}=b+\varepsilon a$. In this
case we seek three commuting flows determined by the conservation law (%
\textbf{\ref{ek}}) written in the potential form%
\begin{equation*}
d\xi =A^{0}dx+A^{1}dt+adt^{1}+bdt^{2}+cdt^{3}.
\end{equation*}%
The corresponding canonical Egorov basic set (\textbf{\ref{thrid}}) (cf. (%
\textbf{\ref{stop}})) in the potential form is%
\begin{equation*}
d\left( 
\begin{array}{c}
\xi ^{1} \\ 
\xi ^{2} \\ 
\xi ^{3}%
\end{array}%
\right) =\left( 
\begin{array}{ccc}
\frac{1}{\varepsilon }\left( \frac{b}{a-c}+a\right) & -\frac{b}{a-c} & \ln
(a-c) \\ 
-\frac{b}{a-c} & b+\varepsilon \frac{b}{a-c} & c-\varepsilon \ln (a-c) \\ 
\ln (a-c) & c-\varepsilon \ln (a-c) & \ln b%
\end{array}%
\right) d\left( 
\begin{array}{c}
t^{1} \\ 
t^{2} \\ 
t^{3}%
\end{array}%
\right) ,
\end{equation*}%
where the Hamiltonian densities are $\mathbf{\tilde{h}}_{1}=a^{2}/2+b\ln
(a-c)$, $\mathbf{\tilde{h}}_{2}=bc-\varepsilon b\ln (a-c)$, $\mathbf{\tilde{h%
}}_{3}=c^{2}/2+b(\ln b-1)+\varepsilon (a-c)[\ln (a-c)-1]$. Every three
components from \textit{any} column of the above $3\times 3$ matrix can be
used as new flat coordinates. Let, for instance, introduce following flat
coordinates%
\begin{equation*}
a_{1}=\frac{1}{\varepsilon }\left( \frac{b}{a-c}+a\right) =\frac{1}{%
\varepsilon }a^{1}\text{, \ \ \ \ \ \ }a_{2}=-\frac{b}{a-c}=a^{3}\text{, \ \
\ \ \ \ \ }a_{3}=\ln (a-c)=a^{2}.
\end{equation*}%
Then taking in account (\textbf{\ref{zima}})%
\begin{equation}
\partial _{t^{k}}a_{n}=\partial _{t^{1}}\left( \frac{\partial ^{2}F}{%
\partial a^{n}\partial a^{k}}\right)  \label{muk}
\end{equation}%
the corresponding solution of the WDVV equation can be found (by two
quadratures)%
\begin{equation*}
F=\frac{(a^{1})^{3}}{6\varepsilon }+a^{1}a^{2}a^{3}-a^{3}e^{a^{2}}+\frac{1}{2%
}a^{2}(a^{3})^{2}-\frac{\varepsilon }{2}a^{3}(a^{2})^{2}+\frac{1}{2}%
(a^{3})^{2}\left( \ln a^{3}-\frac{3}{2}\right) .
\end{equation*}

\textbf{3}. The second degenerate case (the Kodama hydrodynamic reduction (%
\textbf{\ref{kodama}}))%
\begin{equation*}
\lambda =\mu +\frac{c}{\mu -a}+\frac{b^{2}}{2(\mu -a)^{2}}
\end{equation*}%
can be obtained from the previous waterbag hydrodynamic reduction by a
merging of three singular points. The corresponding Egorov hydrodynamic type
system%
\begin{equation*}
a_{t}=\partial _{x}\left( \frac{a^{2}}{2}+c\right) \text{, \ \ \ \ \ }%
b_{t}=\partial _{x}(ab)\text{, \ \ \ \ \ \ }c_{t}=\partial _{x}\left( ac+%
\frac{b^{2}}{2}\right)
\end{equation*}%
possesses the local Hamiltonian structure%
\begin{equation*}
a_{t}=\partial _{x}\frac{\partial \mathbf{h}_{2}}{\partial c}\text{, \ \ \ \
\ \ \ }b_{t}=\partial _{x}\frac{\partial \mathbf{h}_{2}}{\partial b}\text{,
\ \ \ \ \ \ \ }c_{t}=\partial _{x}\frac{\partial \mathbf{h}_{2}}{\partial a},
\end{equation*}%
and the potential of the Egorov metric is $A^{0}=c$. In this case we seek
just \textit{two} commuting flows determined by the conservation law (%
\textbf{\ref{ek}}) written in the potential form%
\begin{equation*}
d\xi =A^{0}dx+A^{1}dt+adt^{1}+bdt^{2},
\end{equation*}%
where $t^{3}\equiv x$.

The corresponding canonical Egorov basic set (\textbf{\ref{thrid}}) (cf. (%
\textbf{\ref{stop}})) in the potential form is%
\begin{equation*}
d\left( 
\begin{array}{c}
\xi ^{3} \\ 
\xi ^{2} \\ 
\xi ^{1}%
\end{array}%
\right) =\left( 
\begin{array}{ccc}
c & b & a \\ 
b & a-c^{2}/(2b^{2}) & c/b \\ 
a & c/b & \ln b%
\end{array}%
\right) d\left( 
\begin{array}{c}
t^{3} \\ 
t^{2} \\ 
t^{1}%
\end{array}%
\right) ,
\end{equation*}%
where the Hamiltonian densities are $\mathbf{\tilde{h}}_{1}=a^{2}/2+c\ln b$, 
$\mathbf{\tilde{h}}_{2}=ab+c^{2}/(2b)$. Every three components from \textit{%
any} column of the above $3\times 3$ matrix can be used as new flat
coordinates. Let, for instance, introduce following flat coordinates%
\begin{equation*}
a_{1}=c=a^{3}\text{, \ \ \ \ \ \ \ }a_{2}=b=a^{2}\text{, \ \ \ \ \ \ \ \ }%
a_{3}=a=a^{1}.
\end{equation*}%
Then (see (\textbf{\ref{muk}})) the corresponding solution of the WDVV
equation can be found (by two quadratures)%
\begin{equation*}
F=\frac{1}{2}(a^{1})^{2}a^{3}+\frac{1}{2}a^{1}(a^{2})^{2}+\frac{1}{2}%
(a^{3})^{2}\ln a^{2}.
\end{equation*}%
This solution was found in \textbf{\cite{Dubr}} (see also \textbf{\cite%
{Aratyn}}).

\textbf{4}. This case is the waterbag hydrodynamic reduction associated with
the equation of the Riemann surface 
\begin{equation*}
\lambda =\frac{\mu ^{2}}{2}+A^{0}-\underset{m=2}{\overset{3}{\sum }}%
\varepsilon _{k}\ln (\mu -a^{k}).
\end{equation*}%
The corresponding Egorov hydrodynamic type system%
\begin{equation*}
A_{t}^{0}=\partial _{x}\left( \underset{m=2}{\overset{3}{\sum }}\varepsilon
_{k}a^{k}\right) \text{, \ \ \ \ \ \ }a_{t}^{k}=\partial _{x}\left( \frac{%
(a^{k})^{2}}{2}+A^{0}\right) \text{, \ \ }k=2,3
\end{equation*}%
possesses the local Hamiltonian structure%
\begin{equation*}
A_{t}^{0}=\partial _{x}\frac{\partial \mathbf{h}_{3}}{\partial A^{0}}\text{,
\ \ \ \ \ \ \ }a_{t}^{k}=\frac{1}{\varepsilon _{k}}\partial _{x}\frac{%
\partial \mathbf{h}_{3}}{\partial a^{k}}\text{, \ \ }k=2,3,
\end{equation*}%
and the potential of the Egorov metric is $A^{0}$. In this case we seek just 
\textit{two} commuting flows determined by the conservation law (\textbf{\ref%
{ek}}) written in the potential form%
\begin{equation*}
d\xi =A^{0}dx+a^{2}dt^{2}+a^{3}dt^{3},
\end{equation*}%
where $t^{1}\equiv x$.

The corresponding canonical Egorov basic set (\textbf{\ref{thrid}}) (cf. (%
\textbf{\ref{stop}})) in the potential form is%
\begin{equation*}
d\left( 
\begin{array}{c}
\xi ^{1} \\ 
\xi ^{2} \\ 
\xi ^{3}%
\end{array}%
\right) =\left( 
\begin{array}{ccc}
A^{0} & a^{2} & a^{3} \\ 
a^{2} & \frac{A^{0}+(a^{2})^{2}/2}{\varepsilon _{2}}-\frac{\varepsilon _{3}}{%
\varepsilon _{2}}\ln (a^{2}-a^{3}) & \ln (a^{2}-a^{3}) \\ 
a^{3} & \ln (a^{2}-a^{3}) & \frac{A^{0}+(a^{3})^{2}/2}{\varepsilon _{3}}-%
\frac{\varepsilon _{2}}{\varepsilon _{3}}\ln (a^{2}-a^{3})%
\end{array}%
\right) d\left( 
\begin{array}{c}
t^{1} \\ 
t^{2} \\ 
t^{3}%
\end{array}%
\right) ,
\end{equation*}%
where the Hamiltonian densities are $\mathbf{\tilde{h}}_{3}=(a^{2})^{3}/6-%
\varepsilon _{3}(a^{2}-a^{3})[\ln (a^{2}-a^{3})-1]$ and $\mathbf{\tilde{h}}%
_{3}=(a^{3})^{3}/6+\varepsilon _{2}(a^{2}-a^{3})[\ln (a^{2}-a^{3})-1]$.
Every three components from \textit{any} column of the above $3\times 3$
matrix can be used as new flat coordinates. For instance, the solution of
the WDVV equation%
\begin{equation*}
F=\frac{1}{6}(a^{1})^{3}+\frac{a^{1}}{2}[\varepsilon
_{2}(a^{2})^{2}+\varepsilon _{3}(a^{3})^{2}]+\frac{\varepsilon _{2}}{24}%
(a^{2})^{4}+\frac{\varepsilon _{3}}{24}(a^{3})^{4}-\frac{\varepsilon
_{2}\varepsilon _{3}}{2}(a^{2}-a^{3})^{2}\left( \ln (a^{2}-a^{3})-\frac{3}{2}%
\right)
\end{equation*}%
can be found (by two quadratures; see (\textbf{\ref{muk}})), where $%
a^{1}=A^{0}$.

\textbf{5}. This degenerate case (the Kodama hydrodynamic reduction (\textbf{%
\ref{kodama}})) 
\begin{equation*}
\lambda =\frac{\mu ^{2}}{2}+A^{0}+\frac{b}{\mu -c}
\end{equation*}%
can be obtained from the previous waterbag hydrodynamic reduction by a
merging of two singular points. The corresponding Egorov hydrodynamic type
system%
\begin{equation*}
A_{t}^{0}=b_{x}\text{, \ \ \ \ \ \ }b_{t}=\partial _{x}(bc)\text{, \ \ \ \ \
\ }c_{t}=\partial _{x}\left( \frac{c^{2}}{2}+A^{0}\right)
\end{equation*}%
possesses the local Hamiltonian structure%
\begin{equation*}
A_{t}^{0}=\partial _{x}\frac{\partial \mathbf{h}_{3}}{\partial A^{0}}\text{,
\ \ \ \ \ \ \ }b_{t}=\partial _{x}\frac{\partial \mathbf{h}_{3}}{\partial c}%
\text{, \ \ \ \ \ \ \ }c_{t}=\partial _{x}\frac{\partial \mathbf{h}_{3}}{%
\partial b},
\end{equation*}%
and the potential of the Egorov metric is $A^{0}$. In this case we seek just 
\textit{one} commuting flow determined by the conservation law (\textbf{\ref%
{ek}}) written in the potential form%
\begin{equation*}
d\xi =A^{0}dx+A^{1}dt+cdt^{3},
\end{equation*}%
where $t^{1}\equiv x$, $t^{2}\equiv t$, $A^{1}\equiv b$.

The corresponding canonical Egorov basic set (\textbf{\ref{thrid}}) (cf. (%
\textbf{\ref{stop}})) in the potential form is%
\begin{equation*}
d\left( 
\begin{array}{c}
\xi ^{1} \\ 
\xi ^{2} \\ 
\xi ^{3}%
\end{array}%
\right) =\left( 
\begin{array}{ccc}
A^{0} & b & c \\ 
b & bc & A^{0}+c^{2}/2 \\ 
c & A^{0}+c^{2}/2 & \ln b%
\end{array}%
\right) d\left( 
\begin{array}{c}
t^{1} \\ 
t^{2} \\ 
t^{3}%
\end{array}%
\right) ,
\end{equation*}%
where the Hamiltonian density is $\mathbf{\tilde{h}}_{2}=cA^{0}+c^{3}/6+b(%
\ln b-1)$. Every three components from \textit{any} column of the above $%
3\times 3$ matrix can be used as new flat coordinates. Let, for instance,
introduce following flat coordinates%
\begin{equation*}
a_{1}=A^{0}=a^{1}\text{, \ \ \ \ \ \ \ }a_{2}=b=a^{3}\text{, \ \ \ \ \ \ \ \ 
}a_{3}=c=a^{2}.
\end{equation*}%
Then (see (\textbf{\ref{muk}})) the corresponding solution of the WDVV
equation can be found (by two quadratures)%
\begin{equation*}
F=\frac{1}{6}(a^{1})^{3}+a^{1}a^{2}a^{3}+\frac{1}{6}(a^{2})^{3}a^{3}+\frac{1%
}{2}(a^{3})^{2}\left( \ln a^{3}-\frac{3}{2}\right) .
\end{equation*}%
This solution was found in \textbf{\cite{Dubr}} (see also \textbf{\cite%
{Aratyn}} and \textbf{\cite{china}}).

\textbf{6}. This case is the waterbag hydrodynamic reduction associated with
the equation of the Riemann surface 
\begin{equation*}
\lambda =\frac{\mu ^{3}}{3}+A^{0}\mu +A^{1}-\varepsilon \ln (\mu -a).
\end{equation*}%
The corresponding Egorov hydrodynamic type system%
\begin{equation*}
A_{t}^{0}=A_{x}^{1}\text{, \ \ \ \ \ \ }A_{t}^{1}=\partial _{x}\left(
\varepsilon a-\frac{(A^{0})^{2}}{2}\right) \text{, \ \ \ \ \ \ }%
a_{t}=\partial _{x}\left( \frac{a^{2}}{2}+A^{0}\right)
\end{equation*}%
possesses the local Hamiltonian structure%
\begin{equation*}
A_{t}^{0}=\partial _{x}\frac{\partial \mathbf{h}_{4}}{\partial A^{1}}\text{,
\ \ \ \ \ \ \ }A_{t}^{1}=\partial _{x}\frac{\partial \mathbf{h}_{4}}{%
\partial A^{0}}\text{, \ \ \ \ \ \ \ }a_{t}=\frac{1}{\varepsilon }\partial
_{x}\frac{\partial \mathbf{h}_{4}}{\partial a},
\end{equation*}%
and the potential of the Egorov metric is $A^{0}$. In this case we seek just 
\textit{one} commuting flow determined by the conservation law (\textbf{\ref%
{ek}}) written in the potential form%
\begin{equation*}
d\xi =A^{0}dx+A^{1}dt+adt^{3},
\end{equation*}%
where $t^{1}\equiv x$, $t^{2}\equiv t$.

The corresponding canonical Egorov basic set (\textbf{\ref{thrid}}) (cf. (%
\textbf{\ref{stop}})) in the potential form is%
\begin{equation*}
d\left( 
\begin{array}{c}
\xi ^{1} \\ 
\xi ^{2} \\ 
\xi ^{3}%
\end{array}%
\right) =\left( 
\begin{array}{ccc}
A^{0} & A^{1} & a \\ 
A^{1} & \varepsilon a-(A^{0})^{2}/2 & A^{0}+a^{2}/2 \\ 
a & A^{0}+a^{2}/2 & (A^{1}+aA^{0}+a^{3}/3)/\varepsilon%
\end{array}%
\right) d\left( 
\begin{array}{c}
t^{1} \\ 
t^{2} \\ 
t^{3}%
\end{array}%
\right) ,
\end{equation*}%
where the Hamiltonian density is $\mathbf{\tilde{h}}%
_{2}=aA^{1}+a^{2}A^{0}/2+(A^{0})^{2}/2+a^{4}/12$. Every three components
from \textit{any} column of the above $3\times 3$ matrix can be used as new
flat coordinates. Let, for instance, introduce following flat coordinates%
\begin{equation*}
a_{1}=A^{0}=a^{2}\text{, \ \ \ \ \ \ \ }a_{2}=A^{1}=a^{1}\text{, \ \ \ \ \ \
\ \ }a_{3}=a=\frac{1}{\varepsilon }a^{3}.
\end{equation*}%
Then (see (\textbf{\ref{muk}})) the corresponding solution of the WDVV
equation can be found (by two quadratures)%
\begin{equation*}
F=\frac{1}{2}(a^{1})^{2}a^{2}+\frac{1}{2\varepsilon }a^{1}(a^{3})^{2}+\frac{1%
}{2}(a^{2})^{2}a^{3}-\frac{1}{24}(a^{2})^{4}+\frac{1}{6\varepsilon ^{2}}%
a^{2}(a^{3})^{3}+\frac{1}{60\varepsilon ^{4}}(a^{3})^{5}.
\end{equation*}

\textbf{7}. This case is the so-called Lax reduction (dispersionless limit
of the Gelfand--Dikey linear problem; see \textbf{\cite{Dubr}} and \textbf{%
\cite{Krich}}) 
\begin{equation*}
\lambda =\frac{\mu ^{4}}{4}+A^{0}\mu ^{2}+A^{1}\mu +A^{2}+\frac{3}{2}%
(A^{0})^{2}.
\end{equation*}%
The corresponding Egorov hydrodynamic type system%
\begin{equation*}
\partial _{t}h_{0}=\partial _{x}h_{1}\text{, \ \ \ \ \ \ }\partial
_{t}h_{1}=\partial _{x}\left( h_{2}-\frac{(h_{0})^{2}}{2}\right) \text{, \ \
\ \ \ \ }\partial _{t}h_{2}=\partial _{x}(-h_{0}h_{1})
\end{equation*}%
possesses the local Hamiltonian structure%
\begin{equation*}
\partial _{t}h_{0}=\partial _{x}\frac{\partial \mathbf{h}_{5}}{\partial h_{2}%
}\text{, \ \ \ \ \ \ \ }\partial _{t}h_{1}=\partial _{x}\frac{\partial 
\mathbf{h}_{5}}{\partial h_{1}}\text{, \ \ \ \ \ \ \ }\partial
_{t}h_{2}=\partial _{x}\frac{\partial \mathbf{h}_{5}}{\partial h_{0}},
\end{equation*}%
the potential of the Egorov metric is $A^{0}=h_{0}$ and $h_{1}=A^{1}$, $%
h_{2}=A^{2}+(A^{0})^{2}$. In this case we seek just \textit{one} commuting
flow determined by the conservation law (\textbf{\ref{ek}}) written in the
potential form%
\begin{equation*}
d\xi =h_{0}dx+h_{1}dt+h_{2}dt^{3},
\end{equation*}%
where $t^{1}\equiv x$, $t^{2}\equiv t$.

The corresponding canonical Egorov basic set (\textbf{\ref{thrid}}) (cf. (%
\textbf{\ref{stop}})) in the potential form is%
\begin{equation*}
d\left( 
\begin{array}{c}
\xi ^{1} \\ 
\xi ^{2} \\ 
\xi ^{3}%
\end{array}%
\right) =\left( 
\begin{array}{ccc}
h_{0} & h_{1} & h_{2} \\ 
h_{1} & h_{2}-(h_{0})^{2}/2 & -h_{0}h_{1} \\ 
h_{2} & -h_{0}h_{1} & (h_{0})^{3}/3-(h_{1})^{2}/2%
\end{array}%
\right) d\left( 
\begin{array}{c}
t^{1} \\ 
t^{2} \\ 
t^{3}%
\end{array}%
\right) ,
\end{equation*}%
where the Hamiltonian density is $\mathbf{\tilde{h}}%
_{2}=(h_{2})^{2}/2-h_{0}(h_{1})^{2}/2+(h_{0})^{4}/12$. Every three
components from \textit{any} column of the above $3\times 3$ matrix can be
used as new flat coordinates. Let, for instance, introduce following flat
coordinates%
\begin{equation*}
a_{1}=h_{0}=a^{3}\text{, \ \ \ \ \ \ \ }a_{2}=h_{1}=a^{2}\text{, \ \ \ \ \ \
\ \ }a_{3}=h_{2}=a^{1}.
\end{equation*}%
Then (see (\textbf{\ref{muk}})) the corresponding solution of the WDVV
equation can be found (by two quadratures)%
\begin{equation*}
F=\frac{1}{2}(a^{1})^{2}a^{3}+\frac{1}{2}a^{1}(a^{2})^{2}-\frac{1}{4}%
(a^{2})^{2}(a^{3})^{2}+\frac{1}{60}(a^{3})^{5}.
\end{equation*}%
This solution was found in \textbf{\cite{Dubr}}.

\textbf{8}. This case associated with the so-called Schwarz-Christoffel map (%
$\Sigma \varepsilon _{k}\neq 1$)%
\begin{equation*}
\lambda =\left( \mu -\sum \varepsilon _{k}a^{k}\right) \underset{m=1}{%
\overset{3}{\prod }}(\mu -a^{k})^{-\varepsilon _{k}},
\end{equation*}%
is considered in the next section.

\textbf{Remark}: In $N$ component case the number of degenerate curves
increases. For example, if $N=4$, then we have following list of these curves%
\begin{eqnarray*}
\lambda &=&\mu -\underset{m=1}{\overset{4}{\sum }}\varepsilon _{k}\ln (\mu
-a^{k})\text{, \ \ \ \ \ \ \ \ \ \ }\lambda =\mu -\underset{m=1}{\overset{2}{%
\sum }}\varepsilon _{k}\ln (\mu -a^{k})+\frac{b}{\mu -c}, \\
&& \\
\lambda &=&\mu -\varepsilon \ln (\mu -u)+\frac{a}{\mu -c}+\frac{b}{(\mu
-c)^{2}}\text{, \ \ \ \ \ \ \ }\lambda =\mu +\frac{a_{1}}{\mu -c_{1}}+\frac{%
a_{2}}{\mu -c_{2}}, \\
&& \\
\lambda &=&\mu +\frac{a}{\mu -u}+\frac{b}{(\mu -u)^{2}}+\frac{c}{(\mu -u)^{3}%
}\text{, \ \ \ \ \ }\lambda =\frac{\mu ^{2}}{2}+A^{0}-\underset{m=1}{\overset%
{3}{\sum }}\varepsilon _{k}\ln (\mu -a^{k})\text{,} \\
&& \\
\lambda &=&\frac{\mu ^{2}}{2}+A^{0}-\varepsilon \ln (\mu -a)+\frac{b}{\mu -c}%
\text{, \ \ \ \ \ \ \ \ \ }\lambda =\frac{\mu ^{2}}{2}+A^{0}+\frac{a}{\mu -u}%
+\frac{b}{(\mu -u)^{2}}\text{,} \\
&& \\
\lambda &=&\frac{\mu ^{3}}{3}+A^{0}\mu +A^{1}+\frac{a}{\mu -u}\text{, \ \ \
\ \ \ \ \ }\lambda =\frac{\mu ^{4}}{4}+A^{0}\mu ^{2}+A^{1}\mu +A^{2}+\frac{3%
}{2}(A^{0})^{2}-\varepsilon \ln (\mu -a), \\
&& \\
\lambda &=&\frac{\mu ^{3}}{3}+A^{0}\mu +A^{1}-\underset{m=1}{\overset{2}{%
\sum }}\varepsilon _{k}\ln (\mu -a^{k})\text{, \ \ \ \ \ \ \ }\lambda
=\left( \mu -\sum \varepsilon _{k}a^{k}\right) \underset{m=1}{\overset{4}{%
\prod }}(\mu -a^{k})^{-\varepsilon _{k}},
\end{eqnarray*}%
where $\Sigma \varepsilon _{k}\neq 1$. In the particular case $\varepsilon
_{k}\equiv \varepsilon $ this is nothing but the Lax hydrodynamic reduction
written in a factorized form.

\section{Modified Benney hydrodynamic chain}

The modified Benney hydrodynamic chain%
\begin{equation}
B_{t}^{n}=B_{x}^{n+1}+B^{0}B_{x}^{n}+nB^{n}B_{x}^{0}\text{, \ \ \ \ \ }n=0%
\text{, }1\text{, }2\text{, ...}  \label{mbc}
\end{equation}%
is connected with the Benney hydrodynamic chain (\textbf{\ref{bm}}) by an
infinite set of the Miura type transformations $%
A^{k}(B^{0},B^{1},...,B^{k+1})$, which can be derived from comparison of two
Riemann mappings (see (\textbf{\ref{rm}}))%
\begin{equation*}
\lambda =p+B^{0}+\frac{B^{1}}{p}+\frac{B^{2}}{p^{2}}+...=p+B^{0}+\frac{A^{0}%
}{p+B^{0}}+\frac{A^{1}}{(p+B^{0})^{2}}+\frac{A^{2}}{(p+B^{0})^{3}}+...,
\end{equation*}%
where the generating function of the Miura type transformations is given by $%
\mu =p+B^{0}$. Then the Gibbons equation (\textbf{\ref{5}}) reduces to%
\begin{equation}
\lambda _{t}-(p+B^{0})\lambda _{x}=\frac{\partial \lambda }{\partial p}\left[
p_{t}-\partial _{x}\left( \frac{p^{2}}{2}+B^{0}p\right) \right] .
\label{gem}
\end{equation}%
Thus, hydrodynamic reductions of both hydrodynamic chains coincide. By this
reason we consider just two ``flat'' hydrodynamic reductions.

The modified Benney hydrodynamic chain is a particular case of the
Kupershmidt hydrodynamic chain (see (\textbf{\ref{kuper}}) and \textbf{\cite%
{Kuper}}).

\textbf{Lemma \cite{Kuper}}: \textit{The modified Benney hydrodynamic chain}%
\begin{equation}
\partial _{t}B_{(\gamma )}^{n}=\partial _{x}B_{(\gamma )}^{n+1}+B_{(\gamma
)}^{0}\partial _{x}B_{(\gamma )}^{n}+(n+\gamma )B_{(\gamma )}^{n}\partial
_{x}B_{(\gamma )}^{0}\text{, \ \ \ \ \ }n=0\text{, }1\text{, }2\text{, ...}
\label{mbch}
\end{equation}%
\textit{is equivalent} (\textbf{\ref{mbc}}) \textit{under an invertible
point transformation}.

\textbf{Proof}: Substitution the series%
\begin{equation}
\lambda =p^{1-\gamma }+(1-\gamma )\underset{k=0}{\overset{\infty }{\sum }}%
\frac{B_{(\gamma )}^{k}}{p^{k+\gamma }}\equiv \lbrack p+\underset{k=0}{%
\overset{\infty }{\sum }}\frac{B^{k}}{p^{k}}]^{1-\gamma }  \label{rim}
\end{equation}%
in the Gibbons equation (\textbf{\ref{gem}}) yields (\textbf{\ref{mbch}}).

\textbf{Remark}: If $\gamma =1$, then%
\begin{equation*}
\lambda =\ln p+\underset{k=0}{\overset{\infty }{\sum }}\frac{B_{(1)}^{k}}{%
p^{k+1}}\equiv \ln [p+\underset{k=0}{\overset{\infty }{\sum }}\frac{B^{k}}{%
p^{k}}].
\end{equation*}

\textbf{Remark}: $B_{(\gamma )}^{0}\equiv B^{0}$.

\subsection{The first local Hamiltonian structure}

If $\gamma =2$, then the modified Benney hydrodynamic chain%
\begin{equation*}
\partial _{t}B_{(2)}^{n}=\partial _{x}B_{(2)}^{n+1}+B_{(2)}^{0}\partial
_{x}B_{(2)}^{n}+(n+2)B_{(2)}^{n}\partial _{x}B_{(2)}^{0}\text{, \ \ \ \ \ }%
n=0\text{, }1\text{, }2\text{, ...}
\end{equation*}%
contains the hydrodynamic reduction%
\begin{equation}
c_{t}^{i}=\partial _{x}\left( \frac{(c^{i})^{2}}{2}+B^{0}c^{i}\right) \text{%
, \ \ \ \ }i=1,2,...,N,  \label{sima}
\end{equation}%
associated with the equation of the Riemann surface%
\begin{equation}
\lambda =\frac{1}{p}\left( 1+\underset{i=1}{\overset{N}{\sum }}\varepsilon
_{i}c^{i}\right) +\underset{i=1}{\overset{N}{\sum }}\varepsilon _{i}\ln
\left( 1-\frac{c^{i}}{p}\right) ,  \label{fir}
\end{equation}%
where $B_{(2)}^{k}=\Sigma \varepsilon _{i}(c^{i})^{k+2}/(k+2)$. This
hydrodynamic type system allows the local Hamiltonian structure%
\begin{equation}
c_{t}^{i}=\frac{1}{2\varepsilon _{i}}\partial _{x}\frac{\partial \mathbf{h}%
_{0}}{\partial c^{i}}\text{, \ \ \ \ }i=1,2,...,N,  \label{d}
\end{equation}%
where the Hamiltonian density $\mathbf{h}_{0}=B_{(2)}^{1}+(B_{(2)}^{0})^{2}%
\equiv A^{0}$.

Since $A^{0}$ is a potential of the Egorov metric, then the generating
function of commuting flows (\textbf{\ref{coma}}) has the conservation law
(see (\textbf{\ref{egor}}) and (\textbf{\ref{dop}}))%
\begin{equation*}
\partial _{\tau (\zeta )}A^{0}=\partial _{x}p(\zeta ).
\end{equation*}%
Then the generating function of commuting flows (\textbf{\ref{coma}}) (cf. (%
\textbf{\ref{cona}}))%
\begin{equation*}
r_{\tau (\zeta )}^{i}=\frac{p(\zeta )}{p^{i}(p^{i}-p(\zeta ))}r_{x}^{i}\text{%
, \ \ \ \ }i=1,2,...,N
\end{equation*}%
is connected with the generating function of commuting flows and
conservation laws (cf. (\textbf{\ref{son}}))%
\begin{equation}
\partial _{\tau (\zeta )}p(\lambda )=\partial _{x}\ln \frac{%
(1+B_{(2)}^{-1})(p(\lambda )-p(\zeta ))}{p(\lambda )p(\zeta )},  \label{gen}
\end{equation}%
where $B_{(2)}^{-1}=\Sigma \varepsilon _{i}c^{i}$, and the first
conservation law is given by%
\begin{equation*}
\partial _{\tau (\zeta )}B^{0}=\partial _{x}\ln \frac{p(\zeta )}{%
1+B_{(2)}^{-1}}.
\end{equation*}%
$N$ commuting flows connected with the corresponding solution of the WDVV
equation are determined by the special limit in the above three formulas $%
p\rightarrow c^{i}$ and $\partial _{\tau (\zeta )}\rightarrow \partial
_{t^{i}}$. Then the generating function of commuting flows (cf. (\textbf{\ref%
{d}}))%
\begin{equation*}
c_{\tau (\zeta )}^{i}=\partial _{x}\ln \frac{(1+B_{(2)}^{-1})(c^{i}-p(\zeta
))}{c^{i}p(\zeta )}
\end{equation*}%
is determined by the Hamiltonian density (cf. (\textbf{\ref{6}}))%
\begin{equation}
\mathbf{\tilde{h}}=(1+B_{(2)}^{-1})\ln (1+B_{(2)}^{-1})-\ln p+\sum
\varepsilon _{m}c^{m}\ln \frac{c^{m}-p(\zeta )}{c^{m}p(\zeta )}.
\label{serv}
\end{equation}%
The substitution of the Taylor series (\textbf{\ref{bls}}) in (\textbf{\ref%
{fir}}) yields an explicit expression for the first conservation law
densities $h_{i}^{(1)}$%
\begin{equation*}
\varepsilon _{i}\ln h_{i}^{(1)}=\sum \varepsilon _{m}\ln c^{i}-\underset{%
m\neq i}{\sum }\varepsilon _{m}\ln (c^{i}-c^{m})-\frac{1+\sum \varepsilon
_{m}c^{m}}{c^{i}}.
\end{equation*}%
Taking into account the above expression the substitution $p\rightarrow
c^{i} $ in (\textbf{\ref{serv}}) leads to the first auxiliary conservation
law densities $\tilde{h}_{i}^{(1)}$ given by%
\begin{equation*}
\tilde{h}_{i}^{(1)}=(1+B_{(2)}^{-1})\ln (1+B_{(2)}^{-1})-\underset{m\neq i}{%
\sum }\varepsilon _{m}(c^{i}-c^{m})\ln (c^{i}-c^{m})+\sum \varepsilon
_{m}[c^{i}\ln c^{i}-c^{m}\ln c^{m}-c^{m}\ln c^{i}]-\ln c^{i},
\end{equation*}%
which determine $N$ necessary commuting flows (cf. (\textbf{\ref{thrid}}), (%
\textbf{\ref{ek}}) and (\textbf{\ref{stop}}))%
\begin{eqnarray*}
c_{t^{k}}^{i} &=&\partial _{x}\ln \frac{(1+B_{(2)}^{-1})(c^{i}-c^{k})}{%
c^{i}c^{k}}\text{,} \\
&& \\
c_{t^{i}}^{i} &=&\frac{1}{\varepsilon _{i}}\partial _{x}\left[ \varepsilon
_{i}\ln (1+B_{(2)}^{-1})-\underset{m\neq i}{\sum }\varepsilon _{m}\ln
(c^{i}-c^{m})+\left( \underset{m\neq i}{\sum }\varepsilon _{m}-\varepsilon
_{i}\right) \ln c^{i}-\frac{1+\sum \varepsilon _{m}c^{m}}{c^{i}}\right] .
\end{eqnarray*}%
A corresponding solution of the WDVV equation can be found in the same way
as in the previous examples. In such a case one must choose following new
flat coordinates%
\begin{eqnarray*}
a_{1} &=&\frac{1}{\varepsilon _{1}}\left[ \varepsilon _{1}\ln
(1+B_{(2)}^{-1})-\underset{m\neq 1}{\sum }\varepsilon _{m}\ln
(c^{1}-c^{m})+\left( \underset{m\neq 1}{\sum }\varepsilon _{m}-\varepsilon
_{1}\right) \ln c^{1}-\frac{1+\sum \varepsilon _{m}c^{m}}{c^{1}}\right] 
\text{,} \\
&& \\
a_{k} &=&\ln \frac{(1+B_{(2)}^{-1})(c^{1}-c^{k})}{c^{1}c^{k}}.
\end{eqnarray*}%
The canonical Egorov basic set is given by the \textit{same} hydrodynamic
type systems (\textbf{\ref{l}}), where $c^{1}(\mathbf{a})$ and $B_{(2)}^{-1}(%
\mathbf{a})$ are solutions of the algebraic equations%
\begin{eqnarray*}
B_{(2)}^{-1} &=&c^{1}\left[ \varepsilon _{1}+(1+B_{(2)}^{-1})\underset{m\neq
1}{\sum }\frac{\varepsilon _{m}}{1+B_{(2)}^{-1}-e^{a_{m}}c^{1}}\right] , \\
&& \\
\sum \varepsilon _{m}a_{m} &=&-\frac{1+B_{(2)}^{-1}}{c^{1}}+\sum \varepsilon
_{m}\ln \frac{1+B_{(2)}^{-1}}{c^{m}}.
\end{eqnarray*}

Thus, a corresponding solution of the WDVV equation is given by the \textit{%
same} expression (see (\textbf{\ref{j}})).

\subsection{The second Hamiltonian structure}

If $\gamma =1$, then the modified Benney hydrodynamic chain%
\begin{equation*}
\partial _{t}B_{(1)}^{n}=\partial _{x}B_{(1)}^{n+1}+B_{(1)}^{0}\partial
_{x}B_{(1)}^{n}+(n+1)B_{(1)}^{n}\partial _{x}B_{(1)}^{0}\text{, \ \ \ \ \ }%
n=0\text{, }1\text{, }2\text{, ...}
\end{equation*}%
contains the hydrodynamic reduction (\textbf{\ref{sima}}) associated with
the equation of the Riemann surface%
\begin{equation}
\lambda =p\underset{k=1}{\overset{N}{\prod }}(p-c^{k})^{-\varepsilon _{k}},
\label{rif}
\end{equation}%
where $B_{(2)}^{k}=\Sigma \varepsilon _{i}(c^{i})^{k+1}/(k+1)$ and $\Sigma
\varepsilon _{i}=0$. This hydrodynamic type system (\textbf{\ref{sima}})
possesses the local Hamiltonian structure%
\begin{equation*}
c_{t}^{i}=\frac{1}{2}\partial _{x}\left[ \frac{1}{\varepsilon _{i}}\frac{%
\partial \mathbf{h}_{1}}{\partial c^{i}}-\underset{n=1}{\overset{N}{\sum }}%
\frac{\partial \mathbf{h}_{1}}{\partial c^{n}}\right] \text{, \ \ \ \ }%
i=1,2,...,N,
\end{equation*}%
where the Hamiltonian density $\mathbf{h}%
_{1}=B_{(1)}^{2}+2B_{(1)}^{0}B_{(1)}^{1}+2(B_{(1)}^{0})^{3}/3\equiv A^{1}$.
Above hydrodynamic chain has the second conservation law%
\begin{equation*}
\partial _{t}[B_{(1)}^{1}+(B_{(1)}^{0})^{2}/2]=\partial
_{x}[B_{(1)}^{2}+2B_{(1)}^{0}B_{(1)}^{1}+2(B_{(1)}^{0})^{3}/3].
\end{equation*}%
Thus, the momentum density $\mathbf{h}_{0}=B_{(1)}^{1}+(B_{(1)}^{0})^{2}/2%
\equiv A^{0}$ is a potential of the Egorov metric. The generating function
of commuting flows and conservation laws (cf. (\textbf{\ref{gen}})) is given
by%
\begin{equation*}
\partial _{\tau (\zeta )}p(\lambda )=\partial _{x}\ln \frac{\prod
(c^{i})^{\varepsilon _{i}}(p(\lambda )-p(\zeta ))}{p(\lambda )p(\zeta )},
\end{equation*}%
where the first conservation law is given by%
\begin{equation*}
\partial _{\tau (\zeta )}B^{0}=\partial _{x}\ln \frac{p(\zeta )}{\prod
(c^{i})^{\varepsilon _{i}}}.
\end{equation*}%
$N$ commuting flows connected with the corresponding solution of the WDVV
equation are determined by the special limit in the above three formulas $%
p\rightarrow c^{i}$ and $\partial _{\tau (\zeta )}\rightarrow \partial
_{t^{i}}$. Then the generating function of commuting flows (cf. (\textbf{\ref%
{d}}))%
\begin{equation*}
c_{\tau (\zeta )}^{i}=\partial _{x}\ln \frac{\prod (c^{n})^{\varepsilon
_{n}}(c^{i}-p(\zeta ))}{c^{i}p(\zeta )}
\end{equation*}%
is determined by the Hamiltonian density (cf. (\textbf{\ref{6}}))%
\begin{equation}
\mathbf{\tilde{h}}=\sum \varepsilon _{m}c^{m}\left[ \sum \varepsilon _{n}\ln
(c^{n}-p)+\ln \frac{c^{m}-p(\zeta )}{c^{m}p(\zeta )}\right] .  \label{sera}
\end{equation}%
The substitution of the Taylor series (\textbf{\ref{bls}}) in (\textbf{\ref%
{rif}}) yields an explicit expression for the first conservation law
densities $h_{i}^{(1)}$%
\begin{equation*}
h_{i}^{(1)}=(c^{i})^{1/\varepsilon _{i}}\underset{m\neq i}{\prod }%
(c^{i}-c^{m})^{-\varepsilon _{m}/\varepsilon _{i}}.
\end{equation*}%
Taking into account the above expression the substitution $p\rightarrow
c^{i} $ in (\textbf{\ref{sera}}) leads to the first auxiliary conservation
law densities $\tilde{h}_{i}^{(1)}$ given by%
\begin{equation*}
\tilde{h}_{i}^{(1)}=-\underset{m\neq i}{\sum }\varepsilon
_{m}(c^{i}-c^{m})\ln (c^{i}-c^{m})-\underset{m\neq i}{\sum }\varepsilon
_{m}c^{m}\ln c^{m}+(1-\varepsilon _{i})c^{i}\ln c^{i},
\end{equation*}%
which determine $N$ necessary commuting flows (cf. (\textbf{\ref{thrid}}), (%
\textbf{\ref{ek}}) and (\textbf{\ref{stop}}))%
\begin{eqnarray*}
c_{t^{k}}^{i} &=&\partial _{x}\ln \frac{\prod (c^{m})^{\varepsilon
_{m}}(c^{i}-c^{k})}{c^{i}c^{k}}\text{,} \\
&& \\
c_{t^{i}}^{i} &=&\frac{1}{\varepsilon _{i}}\partial _{x}\left[ -\underset{%
m\neq i}{\sum }\varepsilon _{m}\ln (c^{i}-c^{m})+\varepsilon _{i}\underset{%
m\neq i}{\sum }\varepsilon _{m}\ln c^{m}+(\varepsilon _{i}-1)^{2}\ln c^{i}%
\right] .
\end{eqnarray*}%
A corresponding solution of the WDVV equation can be found in the same way
as in the previous examples. In such a case one must choose following new
flat coordinates%
\begin{eqnarray*}
a_{1} &=&\frac{1}{\varepsilon _{1}}\left[ -\underset{m\neq 1}{\sum }%
\varepsilon _{m}\ln (c^{1}-c^{m})+\varepsilon _{1}\underset{m\neq 1}{\sum }%
\varepsilon _{m}\ln c^{m}+(\varepsilon _{1}-1)^{2}\ln c^{1}\right] \text{,}
\\
&& \\
a_{k} &=&\ln \frac{\prod (c^{m})^{\varepsilon _{m}}(c^{1}-c^{k})}{c^{1}c^{k}}%
.
\end{eqnarray*}%
Since the above transformation is invertible%
\begin{equation*}
c^{k}=\frac{c^{1}}{1+\exp [a_{k}+\sum \varepsilon _{m}a_{m}]}\text{, \ \ \ \
\ \ }\ln c^{1}=\sum \varepsilon _{m}a_{m}-\underset{m\neq 1}{\sum }%
\varepsilon _{m}\ln \left( 1+\exp [a_{m}+\sum \varepsilon _{n}a_{n}]\right) ,
\end{equation*}%
then the canonical Egorov basic set is given by (cf. (\textbf{\ref{l}}))%
\begin{eqnarray*}
\partial _{t^{n}}a_{1} &=&\partial _{t^{1}}a_{n}\text{, \ \ \ \ \ \ \ }%
\partial _{t^{n}}a_{k}=\partial _{t^{1}}\ln (e^{a_{k}}-e^{a_{n}})\text{, \ \
\ \ }k\neq 1,n, \\
&& \\
\partial _{t^{n}}a_{n} &=&\frac{1}{\varepsilon _{n}}\partial _{t^{1}}\left[
\sum \varepsilon _{m}a_{m}-\varepsilon _{1}a_{n}-\ln \left( 1+e^{a_{n}+\sum
\varepsilon _{m}a_{m}}\right) -\underset{m\neq 1,n}{\sum }\varepsilon
_{m}\ln (e^{a_{m}}-e^{a_{n}})\right] .
\end{eqnarray*}%
Taking into account $a_{k}=a^{k}/\varepsilon _{k}-\Sigma a^{m}$ and $%
a^{k}=\varepsilon _{k}(a_{k}+\Sigma \varepsilon _{m}a_{m})$ the
corresponding solution of the WDVV equation (\textbf{\ref{new}}) can be
found in quadratures (the last summation does not contain the index $1$)%
\begin{equation*}
F=\frac{1-\varepsilon _{1}}{6\varepsilon _{1}}(a^{1})^{3}-\frac{(a^{1})^{2}}{%
2}\underset{m\neq 1}{\sum }a^{m}+\frac{a^{1}}{2}\left[ \underset{m\neq 1}{%
\sum }\frac{(a^{m})^{2}}{\varepsilon _{m}}-\left( \underset{m\neq 1}{\sum }%
a^{m}\right) ^{2}\right] +P_{3}(\mathbf{a})
\end{equation*}%
\begin{equation*}
+\frac{1}{2}\underset{m\neq 1}{\sum }\varepsilon _{m}\left[ \text{Li}%
_{3}\left( e^{a^{m}/\varepsilon _{m}}\right) +\text{Li}_{3}\left(
e^{-a^{m}/\varepsilon _{m}}\right) \right] +\frac{1}{2}\underset{m<k}{\sum }%
\varepsilon _{k}\varepsilon _{m}\left[ \text{Li}_{3}\left(
e^{a^{k}/\varepsilon _{k}-a^{m}/\varepsilon _{m}}\right) +\text{Li}%
_{3}\left( e^{a^{m}/\varepsilon _{m}-a^{k}/\varepsilon _{k}}\right) \right] ,
\end{equation*}%
where%
\begin{equation*}
P_{3}(\mathbf{a})=-\frac{1}{6}\left( \underset{m\neq 1}{\sum }a^{m}\right)
^{3}+\underset{m\neq 1}{\sum }\frac{3\varepsilon _{m}-\varepsilon _{1}-1}{%
24\varepsilon _{m}^{2}}(a^{m})^{3}+\underset{n\neq 1}{\sum }a^{n}\underset{%
m\neq 1}{\sum }\frac{(a^{m})^{2}}{4\varepsilon _{m}}-\underset{m\neq 1}{\sum 
}\frac{(a^{m})^{3}}{4\varepsilon _{m}}.
\end{equation*}

\section{Dispersionless limit of the dBKP/Veselov--Novikov hierarchy}

Under the invertible transformations%
\begin{equation*}
B_{(\gamma )}^{0}=B^{0}\text{, \ \ \ }B_{(\gamma )}^{1}=B^{1}-\frac{\gamma }{%
2}(B^{0})^{2}\text{,\ \ \ \ }B_{(\gamma )}^{2}=B^{2}-\gamma B^{0}B^{1}+\frac{%
\gamma (\gamma +1)}{6}(B^{0})^{3},...
\end{equation*}%
the Kupershmidt hydrodynamic chains (\textbf{\ref{kuper}})%
\begin{equation*}
\partial _{t}B_{(\gamma )}^{k}=\partial _{x}B_{(\gamma )}^{k+1}+\frac{1}{%
\beta }B^{0}\partial _{x}B_{(\gamma )}^{k}+(k+\gamma )B_{(\gamma
)}^{k}B_{x}^{0}\text{, \ \ \ }k=0,1,2,...
\end{equation*}%
are \textit{equivalent} to each other for the \textit{fixed} index $\beta $
and for an arbitrary index $\gamma $. These invertible transformations $%
B_{(\gamma )}^{k}=B_{(\gamma )}^{k}(B^{0},B^{1},...,B^{k})$ can be obtained
by a comparison two Riemann mappings (see \textbf{\cite{Maks+Kuper}})%
\begin{equation*}
\lambda =q+\underset{k=0}{\overset{\infty }{\sum }}\frac{B^{k}}{q^{k}}=\left[
q^{1-\gamma }+(1-\gamma )\underset{k=0}{\overset{\infty }{\sum }}\frac{%
B_{(\gamma )}^{k}}{q^{k+\gamma }}\right] ^{\frac{1}{1-\gamma }}.
\end{equation*}%
If $\gamma =1$, then the equation of the Riemann mapping (\textbf{\ref{rim}}%
) reduces to%
\begin{equation*}
\lambda =\ln q+\underset{k=0}{\overset{\infty }{\sum }}\frac{B_{(1)}^{k}}{%
q^{k+1}}\text{ \ \ \ \ }\Leftrightarrow \text{ \ \ \ \ }\lambda =q\exp
\left( \underset{k=0}{\overset{\infty }{\sum }}\frac{B_{(1)}^{k}}{q^{k+1}}%
\right) .
\end{equation*}%
These above formulas are equivalent up to scaling $\lambda \rightarrow \exp
\lambda $; since the Gibbons equation (where $q=p^{\beta }$; see \textbf{%
\cite{Maks+Kuper}})%
\begin{equation*}
\lambda _{t}-\left( p^{\beta }+\frac{B^{0}}{\beta }\right) \lambda _{x}=%
\frac{\partial \lambda }{\partial p}\left[ p_{t}-\partial _{x}\left( \frac{%
p^{\beta +1}}{\beta +1}+\frac{B^{0}}{\beta }p\right) \right]
\end{equation*}%
is a \textit{linear} equation with respect to $\lambda $, any scaling $%
\lambda \rightarrow \tilde{\lambda}(\lambda )$ is admissible. The
Kupershmidt hydrodynamic chains possess infinitely many local Hamiltonian
structures and hydrodynamic reductions parameterized (in general case) by
the hypergeometric function (see \textbf{\cite{Maks+Kuper}}). The
Kupershmidt hydrodynamic chains possess two component Egorov hydrodynamic
reductions (the ideal gas dynamics, see \textbf{\cite{Manasa}} and \textbf{%
\cite{Maks+Kuper}}). In $N$ component case the Kupershmidt hydrodynamic
chains possess the Egorov hydrodynamic reductions if $\beta =1,2,\infty $
only.

\textbf{Egorov (internal) criterion \cite{Darboux}}: \textit{The Egorov
curvilinear coordinate net is determined by}%
\begin{equation}
\beta _{ik}\beta _{kj}\beta _{ji}=\beta _{ij}\beta _{jk}\beta _{ki}\text{, \
\ \ \ \ \ }i\neq j\neq k.  \label{egorov}
\end{equation}%
Hydrodynamic reductions%
\begin{equation*}
r_{t}^{i}=(q^{i}+\frac{B^{0}}{\beta })r_{x}^{i}\text{, \ \ \ \ \ \ }%
i=1,2,...,N
\end{equation*}%
of the Kupershmidt hydrodynamic chain are described by the Gibbons--Tsarev
system (see \textbf{\cite{Manasa}})%
\begin{equation}
\partial _{i}q^{k}=q^{k}\frac{\partial _{i}B^{0}}{q^{i}-q^{k}}\text{, \ \ \
\ \ }\partial _{ik}B^{0}=\frac{(q^{i}+q^{k})\partial _{i}B^{0}\partial
_{k}B^{0}}{(q^{i}-q^{k})^{2}}\text{, \ \ \ \ \ }i\neq k  \label{mod}
\end{equation}%
reducible (see \textbf{\cite{Maks+Kuper}}) to the canonical form (see 
\textbf{\cite{Gib+Tsar}} and (\textbf{\ref{rm}}), (\textbf{\ref{5}}), (%
\textbf{\ref{zaks}}), (\textbf{\ref{reda}}))%
\begin{equation}
\partial _{i}\mu ^{k}=\frac{\partial _{i}A^{0}}{\mu ^{i}-\mu ^{k}}\text{, \
\ \ \ \ \ }\partial _{ik}A^{0}=\frac{\partial _{i}A^{0}\partial _{k}A^{0}}{%
(\mu ^{i}-\mu ^{k})^{2}}\text{, \ \ \ \ \ }i\neq k  \label{start}
\end{equation}%
by the transformation%
\begin{equation*}
\mu ^{i}=q^{i}+B^{0}\text{, \ \ \ \ \ \ }\partial _{i}A^{0}=q^{i}\partial
_{i}B^{0}.
\end{equation*}%
It is easy to check that hydrodynamic reductions (\textbf{\ref{reda}}) are
the Egorov hydrodynamic reductions. Indeed, taking into account the
Combescure transformation (see \textbf{\cite{Tsar}}) $\tilde{H}_{i}=\mu
^{i}H_{i}$, where $\tilde{H}_{i}$ and $H_{i}$ are different solutions of the
linear problem (\textbf{\ref{lin}}), the Gibbons--Tsarev system (\textbf{\ref%
{start}}) can be written in the form%
\begin{equation*}
\beta _{ik}=\frac{H_{i}^{3}H_{k}^{3}}{(\tilde{H}_{k}H_{i}-H_{k}\tilde{H}%
_{i})^{2}}=\frac{\partial _{i}\tilde{H}_{k}}{\tilde{H}_{i}}=\frac{\partial
_{i}H_{k}}{H_{i}}\text{, \ \ \ \ \ }i\neq k,
\end{equation*}%
where $\partial _{i}A^{0}=H_{i}^{2}$. It means, that the rotation
coefficients $\beta _{ik}$ are symmetric. However, in general case this is
no longer true for an arbitrary index $\beta $. Indeed, the Gibbons--Tsarev
system (\textbf{\ref{mod}}) can be written via Lame coefficients and
rotation coefficients of the curvilinear coordinate nets. Introducing%
\begin{equation*}
\mu ^{i}+B^{0}=\frac{\tilde{H}_{i}}{H_{i}}\text{, \ \ \ \ \ }\beta _{ik}=%
\frac{\partial _{i}\tilde{H}_{k}}{\tilde{H}_{i}}=\frac{\partial _{i}H_{k}}{%
H_{i}}\text{, \ \ \ \ \ \ }i\neq k,
\end{equation*}%
one can obtain the Gibbons--Tsarev system written in another form%
\begin{equation*}
\beta _{ik}=\frac{\psi _{i}H_{i}H_{k}^{2}[\tilde{H}_{i}H_{k}+(\beta -1)H_{i}%
\tilde{H}_{k}-B_{0}H_{i}H_{k}]}{\beta (\tilde{H}_{i}H_{k}-H_{i}\tilde{H}%
_{k})^{2}}=\frac{\partial _{i}\tilde{H}_{k}}{\tilde{H}_{i}}=\frac{\partial
_{i}H_{k}}{H_{i}}=\frac{\partial _{k}\psi _{i}}{\psi _{k}}\text{,\ \ \ }%
i\neq k,
\end{equation*}%
where $\partial _{i}B_{0}=\psi _{i}H_{i}$ and $\psi _{i}$ is a solution of
the adjoint linear problem to (\textbf{\ref{lin}}) (see details in \textbf{%
\cite{Tsar}}). These rotation coefficients $\beta _{ik}$ are no symmetric in
general case. However, substituting the above expression in (\textbf{\ref%
{egorov}}) one can obtain three exceptional cases $\beta =1$ (the modified
Benney hydrodynamic chain), $\beta =2$ (the hydrodynamic chain associated
with the dBKP/Veselov--Novikov hierarchy; see \textbf{\cite{Bogdan}}, see 
\textbf{\cite{Chen+Tu}}) and $\beta =\infty $ (continuum limit of 2DToda
Lattice).

\subsection{Waterbag hydrodynamic reduction}

In this section we restrict our consideration on the waterbag hydrodynamic
reductions for $\beta =2$. The first hydrodynamic chain of the
dBKP/Veselov--Novikov hierarchy 
\begin{equation*}
\partial _{t}B_{(1)}^{k}=\partial _{x}B_{(1)}^{k+1}+\frac{1}{2}B^{0}\partial
_{x}B_{(1)}^{k}+(k+1)B_{(1)}^{k}B_{x}^{0}\text{, \ \ \ }k=0,1,2,...
\end{equation*}%
contains the waterbag hydrodynamic reduction%
\begin{equation*}
c_{t}^{i}=\partial _{x}\left( \frac{(c^{i})^{3}}{2}+C^{0}c^{i}\right) \text{%
, \ \ \ \ }i=1,2,...,N,
\end{equation*}%
associated with the equation of the Riemann surface%
\begin{equation*}
\lambda =2\ln p-\underset{i=1}{\overset{N}{\sum }}\varepsilon _{i}\ln \left(
1-\frac{(c^{i})^{2}}{p^{2}}\right) ,
\end{equation*}%
where $B_{(1)}^{k}=\Sigma \varepsilon _{i}(c^{i})^{2k+2}/(k+1)$ and $%
C^{0}=B^{0}/2$. This hydrodynamic type system allows the local Hamiltonian
structure%
\begin{equation*}
c_{t}^{i}=\frac{1}{6\varepsilon _{i}}\partial _{x}\frac{\partial \mathbf{h}%
_{0}}{\partial c^{i}}\text{, \ \ \ \ }i=1,2,...,N,
\end{equation*}%
where the Hamiltonian density $\mathbf{h}%
_{0}=B_{(1)}^{1}+3(B_{(1)}^{0})^{2}/4$. Since this hydrodynamic chain has
the couple of conservation laws (\textbf{\ref{egor}})%
\begin{equation*}
B_{t}^{0}=\partial _{x}\left( B_{(1)}^{1}+\frac{3}{4}(B_{(1)}^{0})^{2}%
\right) \text{,\ \ \ \ \ \ }\partial _{t}\left( B_{(1)}^{1}+\frac{3}{4}%
(B_{(1)}^{0})^{2}\right) =\partial _{x}\left(
B_{(1)}^{2}+2B_{(1)}^{0}B_{(1)}^{1}+\frac{3}{4}(B_{(1)}^{0})^{3}\right) ,
\end{equation*}%
then $C^{0}$ is a potential of the Egorov metric. Then the generating
function of commuting flows (\textbf{\ref{coma}}) has the conservation law
(see (\textbf{\ref{egor}}) and (\textbf{\ref{dop}}))%
\begin{equation*}
\partial _{\tau (\zeta )}C^{0}=\partial _{x}p(\zeta ).
\end{equation*}%
Thus, the generating function of conservation laws and commuting flows is
given by (cf. (\textbf{\ref{son}}) and (\textbf{\ref{gen}}); see, for
instance, \textbf{\cite{Bogdan}}, \textbf{\cite{Chen+Tu}}) and \textbf{\cite%
{Maks+Kuper}})%
\begin{equation*}
\partial _{\tau (\zeta )}p(\lambda )=\frac{1}{2}\partial _{x}\ln \frac{%
p(\lambda )-p(\zeta )}{p(\lambda )+p(\zeta )}.
\end{equation*}%
$N$ commuting flows connected with the corresponding solution of the WDVV
equation are determined by the special limit in the above three formulas $%
p\rightarrow c^{i}$ and $\partial _{\tau (\zeta )}\rightarrow \partial
_{t^{i}}$. Then the generating function of commuting flows (cf. (\textbf{\ref%
{d}}))%
\begin{equation*}
c_{\tau (\zeta )}^{i}=\frac{1}{2}\partial _{x}\ln \frac{c^{i}-p(\zeta )}{%
c^{i}+p(\zeta )}
\end{equation*}%
is determined by the Hamiltonian density (cf. (\textbf{\ref{6}}))%
\begin{equation*}
\mathbf{\tilde{h}}=\frac{1}{2}\sum \varepsilon _{m}c^{m}\ln \frac{%
c^{m}-p(\zeta )}{c^{m}+p(\zeta )}.
\end{equation*}%
The substitution of the Taylor series (\textbf{\ref{bls}}) in (\textbf{\ref%
{rif}}) yields an explicit expression for the first conservation law
densities $h_{i}^{(1)}$%
\begin{equation*}
h_{i}^{(1)}=\frac{1}{2}(c^{i})^{2(1+\Sigma \varepsilon _{m})/\varepsilon
_{i}-1}\underset{m\neq i}{\prod }\left[ (c^{i})^{2}-(c^{m})^{2}\right]
^{-\varepsilon _{m}/\varepsilon _{i}}.
\end{equation*}%
Taking into account the above expression the substitution $p\rightarrow
c^{i} $ in (\textbf{\ref{sera}}) leads to the first auxiliary conservation
law densities $\tilde{h}_{i}^{(1)}$ given by%
\begin{equation*}
\tilde{h}_{i}^{(1)}=-\frac{1}{2}\underset{m\neq i}{\sum }\varepsilon
_{m}(c^{i}-c^{m})\ln (c^{i}-c^{m})-\frac{1}{2}\underset{m\neq i}{\sum }%
\varepsilon _{m}(c^{i}+c^{m})\ln (c^{i}+c^{m})+\left( 1+\underset{m\neq i}{%
\sum }\varepsilon _{m}\right) c^{i}\ln c^{i},
\end{equation*}%
which determine $N$ necessary commuting flows (cf. (\textbf{\ref{thrid}}), (%
\textbf{\ref{ek}}) and (\textbf{\ref{stop}}))%
\begin{eqnarray*}
c_{t^{k}}^{i} &=&\frac{1}{2}\partial _{x}\ln \frac{c^{i}-c^{k}}{c^{i}+c^{k}}%
\text{,} \\
&& \\
c_{t^{i}}^{i} &=&\frac{1}{\varepsilon _{i}}\partial _{x}\left[ \left( 1+%
\underset{m\neq i}{\sum }\varepsilon _{m}\right) \ln c^{i}-\frac{1}{2}%
\underset{m\neq i}{\sum }\varepsilon _{m}\ln [(c^{i})^{2}-(c^{m})^{2}]\right]
.
\end{eqnarray*}%
A corresponding solution of the WDVV equation can be found in the same way
as in the previous examples. In such a case one must choose following new
flat coordinates%
\begin{eqnarray*}
a_{1} &=&\frac{1}{\varepsilon _{1}}\left[ \left( 1+\underset{m\neq 1}{\sum }%
\varepsilon _{m}\right) \ln c^{1}-\frac{1}{2}\underset{m\neq 1}{\sum }%
\varepsilon _{m}\ln [(c^{1})^{2}-(c^{m})^{2}]\right] \text{,} \\
&& \\
a_{k} &=&\frac{1}{2}\partial _{x}\ln \frac{c^{1}-c^{k}}{c^{1}+c^{k}}.
\end{eqnarray*}%
Since the above transformation is invertible%
\begin{equation*}
c^{k}=-c^{1}\tan a_{k}\text{, \ \ \ \ \ \ }\ln c^{1}=\varepsilon _{1}a_{1}-%
\underset{m\neq 1}{\sum }\varepsilon _{m}\ln \cosh a_{m},
\end{equation*}%
then the canonical Egorov basic set is given by (cf. (\textbf{\ref{l}}))%
\begin{eqnarray*}
\partial _{t^{n}}a_{1} &=&\partial _{t^{1}}a_{n}\text{, \ \ \ \ \ \ \ }%
\partial _{t^{n}}a_{k}=\frac{1}{2}\partial _{t^{1}}\ln \frac{\sinh
(a_{k}-a_{n})}{\sinh (a_{k}+a_{n})}\text{, \ \ \ \ }k\neq 1,n, \\
&& \\
\partial _{t^{n}}a_{n} &=&\frac{1}{\varepsilon _{n}}\partial _{t^{1}}\left[
\varepsilon _{1}a_{1}+\left( 1+\underset{m\neq n}{\sum }\varepsilon
_{m}\right) \ln \tan a_{n}+(\varepsilon _{1}-\varepsilon _{n})\ln \cosh
a_{n}-\frac{1}{2}\underset{m\neq 1,n}{\sum }\varepsilon _{m}A_{mn}\right] ,
\end{eqnarray*}%
where%
\begin{equation*}
A_{mn}=\ln (\tanh ^{2}a_{m}-\tanh ^{2}a_{n})+2\ln \cosh a_{m}.
\end{equation*}%
The corresponding solution of the WDVV equation (\textbf{\ref{new}}) can be
found in quadratures%
\begin{equation*}
F=\frac{\varepsilon _{1}^{2}}{6}(a_{1})^{3}+\frac{\varepsilon _{1}a_{1}}{2}%
\underset{m\neq 1}{\sum }\varepsilon _{m}(a_{m})^{2}
\end{equation*}%
\begin{eqnarray*}
&&+\underset{m\neq 1}{\sum }\frac{\varepsilon _{m}(1+\varepsilon _{m})}{32}[%
\text{Li}_{3}(e^{4a_{m}})+\text{Li}_{3}(e^{-4a_{m}})]-\frac{2+\sum
\varepsilon _{n}}{8}\underset{m\neq 1}{\sum }\varepsilon _{m}\left[ \text{Li}%
_{3}\left( e^{2a_{m}}\right) +\text{Li}_{3}\left( e^{-2a_{m}}\right) \right]
\\
&& \\
&&+\frac{1}{16}\underset{m<k}{\sum }\varepsilon _{k}\varepsilon _{m}[\text{Li%
}_{3}\left( e^{2a_{k}-2a_{m}}\right) +\text{Li}_{3}\left(
e^{2a_{m}-2a_{k}}\right) +\text{Li}_{3}\left( e^{2a_{k}+2a_{m}}\right) +%
\text{Li}_{3}\left( e^{-2a_{m}-2a_{k}}\right) ],
\end{eqnarray*}%
where $a^{k}=\varepsilon _{k}a_{k}$ (the last summation does not contain the
index $1$).

\subsection{Modified dBKP/Veselov--Novikov hierarchy}

The simplest hydrodynamic chain in the classification of the Egorov
integrable hydrodynamic chains (see \textbf{\cite{Maks+Egor}}) is associated
with 2+1 nonlinear equation%
\begin{equation*}
\Omega _{tt}=\Omega _{xy}+\frac{1}{2}\Omega _{xt}^{2}-\exp (-2\Omega _{xx})
\end{equation*}%
generalizing corresponding 2+1 nonlinear equations from the dKP, dBKP and
2dDTL hierarchies (see details in \textbf{\cite{Maks+Egor}}). However, also
this equation is associated with the modified dBKP/Veselov--Novikov
hierarchy (details will be published elsewhere). Nevertheless, in this
sub-section we derive corresponding solution of the WDVV equation (\textbf{%
\ref{new}}) independently. To avoid repetition of the above similar
computation, let us briefly (see corresponding details in \textbf{\cite%
{Maks+Egor}}) describe all necessary actions.

\textbf{1}. The simplest hydrodynamic reductions of the above 2+1 nonlinear
equation are given by (see \textbf{\cite{Maks+algebr}} and \textbf{\cite%
{Maks+Hamch}})%
\begin{equation*}
c_{t}^{i}=\partial _{x}\left[ e^{-\mathbf{h}_{0}}\left(
e^{c^{i}}-e^{-c^{i}}\right) \right] ,
\end{equation*}%
where the generating function of conservation laws is given by (see \textbf{%
\cite{Maks+Egor}})%
\begin{equation*}
p_{t}=\partial _{x}\left[ e^{-\mathbf{h}_{0}}\left( e^{p}-e^{-p}\right) %
\right] .
\end{equation*}

\textbf{2}. The above hydrodynamic type system has the local Hamiltonian
structure (see \textbf{\cite{Maks+Hamch}})%
\begin{equation*}
c_{t}^{i}=\frac{1}{\varepsilon _{i}}\partial _{x}\frac{\partial \mathbf{h}}{%
\partial c^{i}}
\end{equation*}%
iff $\mathbf{h}\equiv \mathbf{h}_{0}=\ln (\Sigma \varepsilon _{m}\cosh
c^{m}) $.

\textbf{3}. The generating function of commuting flows and conservation laws
is given by (see \textbf{\cite{Maks+Egor}})%
\begin{equation}
\partial _{\tau (\zeta )}p(\lambda )=\partial _{x}\ln \frac{e^{p(\lambda
)+p(\zeta )}-1}{e^{p(\lambda )}-e^{p(\zeta )}}.  \label{vn}
\end{equation}

\textbf{4}. The equation of the Riemann surface for the above hydrodynamic
reduction is given by (see \textbf{\cite{Maks+algebr}})%
\begin{equation*}
\lambda =\sum \varepsilon _{m}\ln \frac{\cosh p-\cosh c^{m}}{\sinh p}.
\end{equation*}

\textbf{5}. Let us consider conservation law densities at the vicinity of
each puncture $p^{(i)}=c^{i}$ (i.e. $p^{(i)}=c^{i}+\tilde{\lambda}b^{i}+...$%
, where $\tilde{\lambda}(\lambda )$ is a local parameter). Then applying the
B\"{u}rmann--Lagrange series (see \textbf{\cite{Lavr}}) one can compute%
\begin{equation*}
\varepsilon _{i}\ln b^{i}=\underset{m\neq i}{\sum }\varepsilon _{m}\ln \frac{%
\sinh c^{i}}{\cosh c^{i}-\cosh c^{m}}.
\end{equation*}

\textbf{6}. The substitution the above series $p^{(i)}=c^{i}+\tilde{\lambda}%
b^{i}+...$ in (\textbf{\ref{vn}}) leads to the Hamiltonian hydrodynamic type
commuting systems%
\begin{equation*}
c_{t^{i}}^{i}=\partial _{x}\ln \frac{e^{c^{i}}-e^{-c^{i}}}{b^{i}(\mathbf{c})}%
\text{, \ \ \ \ \ \ }c_{t^{k}}^{i}=\ln \frac{e^{c^{i}+c^{k}}-1}{%
e^{c^{i}}-e^{c^{k}}}\text{, \ \ \ \ }i\neq k.
\end{equation*}

\textbf{7}. Let us introduce a new set of the flat coordinates%
\begin{equation*}
a_{1}=\ln \frac{e^{c^{1}}-e^{-c^{1}}}{b^{1}(\mathbf{c})}\text{, \ \ \ \ \ }%
a_{k}=\ln \frac{e^{c^{1}+c^{k}}-1}{e^{c^{1}}-e^{c^{k}}}\text{, \ \ \ \ }%
i\neq k.
\end{equation*}%
Then the corresponding solution of the WDVV equation is given by

\begin{equation*}
F=\frac{\varepsilon _{1}^{2}}{6}(a_{1})^{3}+\frac{\varepsilon _{1}a_{1}}{2}%
\underset{m\neq 1}{\sum }\varepsilon _{m}(a_{m})^{2}+\underset{m\neq 1}{\sum 
}\frac{\varepsilon _{m}(\Sigma \varepsilon _{n}-2\varepsilon _{m})}{8}[\text{%
Li}_{3}(e^{2a_{m}})+\text{Li}_{3}(e^{-2a_{m}})]
\end{equation*}%
\begin{equation*}
-\frac{1}{2}\underset{m<k}{\sum }\varepsilon _{k}\varepsilon _{m}[\text{Li}%
_{3}\left( e^{a_{k}-a_{m}}\right) +\text{Li}_{3}\left(
e^{a_{m}-a_{k}}\right) +\text{Li}_{3}\left( e^{a_{k}+a_{m}}\right) +\text{Li}%
_{3}\left( e^{-a_{m}-a_{k}}\right) ],
\end{equation*}%
where $a^{k}=\varepsilon _{k}a_{k}$ (the last summation does not contain the
index $1$) and the canonical Egorov basic set is given by (cf. with the
previous example)%
\begin{eqnarray*}
\partial _{t^{n}}a_{1} &=&\partial _{t^{1}}a_{n}\text{, \ \ \ \ \ \ \ }%
\partial _{t^{n}}a_{k}=\partial _{t^{1}}\ln \frac{\sinh (\frac{a_{k}+a_{n}}{2%
})}{\sinh (\frac{a_{k}-a_{n}}{2})}\text{, \ \ \ \ }k\neq 1,n, \\
&& \\
\partial _{t^{n}}a_{n} &=&\frac{1}{\varepsilon _{n}}\partial _{t^{1}}\left[
\varepsilon _{1}a_{1}+\left( \varepsilon _{n}-\underset{m\neq n}{\sum }%
\varepsilon _{m}\right) \ln \sinh a_{n}+\underset{m\neq 1,n}{\sum }%
\varepsilon _{m}\ln \left( \cosh a_{n}-\cosh a_{m}\right) \right] .
\end{eqnarray*}

\section{Conclusion and outlook}

In this paper we consider several explicit solution of the WDVV equation
expressed via elementary and special functions of flat coordinates. These
solutions are connected with the Hamiltonian hydrodynamic reductions of some
Egorov hydrodynamic chains such the Benney hydrodynamic chain. Since the
Benney hydrodynamic chain has infinitely many Hamiltonian hydrodynamic
reductions, then infinitely many corresponding solutions of the WDVV
equation can be found. In the same way more complicated solutions of the
WDVV equation can be constructed by virtue of the Hamiltonian method of
hydrodynamic reductions (see \textbf{\cite{Maks+Hamch}}) extracted from more
complicated Egorov hydrodynamic chains (see \textbf{\cite{Maks+Egor}}).

The theory of the WDVV equation is closely connected with the theory of $N$
orthogonal curvilinear coordinate nets (see \textbf{\cite{Dubr}}, \textbf{%
\cite{Krich}}, \textbf{\cite{Maks+Egor+int}}, \textbf{\cite{Strachan}}, 
\textbf{\cite{Zakh+geom}}). In this paper we are able completely to ignore
computations in Riemann invariants. All computation are made via flat
coordinates.

In the sub-section 4.2 the list of three component solutions of the WDVV
equation associated with corresponding flat hydrodynamic reductions is
presented. However, this list is not complete. Three extra hydrodynamic type
systems are connected with the equations of the Riemann surface%
\begin{equation*}
\lambda =\mu +\frac{b}{\mu -c}-\varepsilon _{1}\ln (\mu -c)-\varepsilon
_{2}\ln (\mu -a)\text{, \ \ \ \ \ \ \ \ }\lambda =\mu +\frac{c}{\mu -a}+%
\frac{b^{2}}{2(\mu -a)^{2}}-\varepsilon \ln (\mu -a),
\end{equation*}%
\begin{equation*}
\lambda =\frac{\mu ^{2}}{2}+A^{0}+\frac{b}{\mu -c}-\varepsilon \ln (\mu -c).
\end{equation*}%
The third case was considered in details in \textbf{\cite{Strachan}}. The
first two cases also are embedded in the framework of ``logarithmic''
deformations of rational curves (see also \textbf{\cite{Strachan}}). If $%
\varepsilon =0$, the first case is Example \textbf{2}, the second case is
Example \textbf{3}, the third case is Example \textbf{5} from the
sub-section 4.2. Thus, $\varepsilon $ is a deformation parameter. Moreover,
the logarithmic deformation ($\varepsilon _{m}$ are arbitrary deformation
parameters) of the Krichever truncation (\textbf{\ref{krich}})%
\begin{equation*}
\lambda =\frac{\mu ^{K+1}}{K+1}+\underset{k=0}{\overset{K-1}{\sum }}%
Q_{K-1-k}(\mathbf{A})\mu ^{k}+\underset{m=1}{\overset{M}{\sum }}\underset{n=1%
}{\overset{N_{m}}{\sum }}\frac{b_{m,n}}{(\mu -c^{m})^{n}}-\underset{m=1}{%
\overset{M}{\sum }}\varepsilon _{m}\ln (\mu -c^{m})
\end{equation*}%
preserves just one of two local Hamiltonian structures associated with
corresponding hydrodynamic reductions. It means, that if $\varepsilon _{m}$
are small parameters, then the above Riemann surfaces are connected with
quasi-Frobenius manifolds, which should be investigated elsewhere.

\section*{Acknowledgement}

I thank Michail Feigin, Eugeni Ferapontov, Luuk Hoevenaars, Yuji Kodama and
Sergey Tsarev for their help and clarifying discussions.

Especially I would like to thank Ian Strachan for sending me his recent
manuscript \textbf{\cite{Strachan}}.

I grateful to the Institute of Mathematics in Taipei (Taiwan) where some
part of this work was made, and especially Derchyi Wu, Jen-Hsu Chang,
Ming-Hien Tu and Jyh-Hao Lee for fruitful discussions.%

\end{document}